\documentclass[a4paper]{article}
\usepackage{a4wide}
\usepackage[UKenglish]{babel}
\usepackage{graphicx,subfigure,caption}
\graphicspath{{./}{figures/}}
\usepackage[colorlinks=true,linkcolor=blue,citecolor=red]{hyperref}
\usepackage{xcolor}
\usepackage{amssymb,amsthm,empheq,bbold}
\usepackage{calrsfs}
	\DeclareMathAlphabet{\pazocal}{OMS}{zplm}{m}{n}
\usepackage[inline]{enumitem}
\usepackage[ruled,vlined,linesnumbered]{algorithm2e}

\newcommand{\abs}[1]{\left\vert#1\right\vert}

\newcommand{\B}{\text{B}}
\newcommand{\cB}{\pazocal{B}}
\newcommand{\eps}{\varepsilon}
\newcommand{\N}{\mathbb{N}}
\newcommand{\pr}[1]{{}^\prime\!#1}
\newcommand{\cQ}{\mathcal{Q}}
\newcommand{\R}{\mathbb{R}}
\newcommand{\cR}{\mathcal{R}}
\newcommand{\supp}[1]{\operatorname{supp}(#1)}

\newcommand{\cV}{\pazocal{V}}

\newcommand{\Z}{\mathbb{Z}}

\newtheorem{assumption}{Assumption}[section]
\theoremstyle{remark}\newtheorem*{remark}{Remark}
\usepackage{authblk}
\allowdisplaybreaks

\title{Macroscopic limits of non-local kinetic descriptions of vehicular traffic}
\author[$\ast$]{F. A. Chiarello}
\author[$\ast\ast$]{A. Tosin}
\affil[$\ast$]{{\footnotesize Dipartimento di Ingegneria e Scienze dell'Informazione e Matematica, University of L'Aquila, Italy}}
\affil[$\ast\ast$]{{\footnotesize Department of Mathematical Sciences ``G. L. Lagrange'', Politecnico di Torino, Italy}}
\date{}
		
\begin{document}
\maketitle
	
\begin{abstract}
We study the derivation of macroscopic traffic models out of optimal speed and  follow-the-leader particle dynamics as hydrodynamic limits of non-local Povzner-type kinetic equations. As a first step, we show that optimal speed vehicle dynamics produce a first order macroscopic model with non-local flux. Next, we show that non-local follow-the-leader vehicle dynamics have a universal macroscopic counterpart in the second order Aw-Rascle-Zhang traffic model, at least when the non-locality of the interactions is sufficiently small. Finally, we show that the same qualitative result holds also for a general class of follow-the-leader dynamics based on the headway of the vehicles rather than on their speed. We also investigate the correspondence between the solutions to particle models and their macroscopic limits by means of numerical simulations.
\medskip


\medskip

\end{abstract}
	
\section{Introduction}
One of the most celebrated macroscopic traffic models based on fluid dynamic equations is the Lighthill-Whitham-Richards (LWR) model~\cite{lighthill1955PRSLA,richards1956OR}, which consists in a scalar equation expressing the conservation of the number of cars on the road:
$$ \partial_t\rho+\partial_x(\rho\cV(\rho))=0, $$
$\rho=\rho(x,t)$ being the mean traffic density, i.e. the number of vehicles per unit length of the road, in the point $x$ at time $t$ and $\cV$ the mean traffic speed. Although widely used in traffic applications thanks to its simplicity and to a well-established analytical theory, this model has some limitations. We recall, in particular, that it allows for speed discontinuities resulting in infinite accelerations. 
Great efforts have been devoted to overcome these drawbacks, among which it is worth mentioning some \textit{non-local} versions of the LWR model proposed in very recent times, see e.g.,~\cite{blandin2016NM,chiarello2018M2AN,friedrich2018NHM,goatin2016NHM,keimer2017JDE}. Most of them share the idea to model the mean traffic speed as a downstream convolution between the traffic density and a prescribed decreasing kernel. The convolution introduces naturally a smooth non-locality in the flux of vehicles, which produces Lipschitz-continuous speeds in $x$ and $t$ ensuring bounded accelerations. Moreover, such a non-locality may describe the anisotropic behaviour of drivers adapting the speed of their vehicles to that of vehicles ahead, caring particularly of close ones. 

Another option to cure the aforesaid limitations of the LWR model is to consider second order macroscopic models, in which the vehicle density $\rho$ and mean speed $u=u(x,t)$ are regarded as two independent hydrodynamic parameters. One of the most popular models in this class is the Aw-Rascle-Zhang (ARZ) model~\cite{aw2000SIAP,zhang2002TRB}, which consists in a system of two scalar equations expressing the conservation of the number of cars on the road and the balance of linear momentum:
$$	\begin{cases}
		\partial_t\rho+\partial_x(\rho u)=0 \\
		\partial_tu+(u-\rho p'(\rho))\partial_xu=0.
	\end{cases} $$
Unlike other second order models, such as e.g.,~\cite{payne1971MMPS}, which were affected by physical inconsistencies due to a too strict link with the fluid dynamic equations inspiring them, cf.~\cite{daganzo1995TR}, the ARZ model accounts correctly for the anticipation ability of the drivers through a prescribed (pseudo-) pressure of traffic $p=p(\rho)$. Notice that such an anticipation ability may be regarded as a non-locality in vehicle interactions.

From this discussion, it is clear that non-local models are motivated by the necessity to provide a description of car flow closer to the actual physics of traffic. On the other hand, their construction relies largely on heuristic ideas. As a matter of fact, the non-local traffic models mentioned above, including the ARZ model, have been also obtained from microscopic descriptions by means of many-particle limits, see e.g.,~\cite{chiarello2020SIAM,difrancesco2017MBE,difrancesco2015ARMA,goatin2017CMS}. In these cases, the adopted procedure consists in showing that the solutions of the macroscopic models can be recovered as limits of the solutions of selected microscopic models used as educated guesses for the discretisation of the former.

The main contribution of this paper is instead to recover first and second order traffic models as \textit{physical limits} of fundamental \textit{non-local particle dynamics}. This way, we will establish \textit{structural} links between microscopic and macroscopic models genuinely grounded on \textit{first principles} instead of simply assessing the consistency of \textit{ad-hoc} particle discretisations of macroscopic models.

To this purpose, we will make use of concepts and tools from collisional \textit{kinetic theory}, which since almost twenty years has been systematically rediscovered as a powerful and flexible mathematical approach to interacting multi-agent systems~\cite{pareschi2013BOOK} often very different from the gas molecules which first inspired the work by Boltzmann. The pioneer of the application of kinetic theory to the modelling of traffic flow was Prigogine~\cite{prigogine1960OR,prigogine1971BOOK}. Years later, other scholars started again to apply it to the mathematical investigation of traffic flow phenomena~\cite{borsche2022PHYSA,herty2003SISC,herty2010KRM,herty2007NHM,herty2020KRM,klar1997JSP,klar2000SIAP}. Nowadays, the literature includes several contributions touching also quite modern applications, such as e.g., driver-assist and autonomous vehicles~\cite{chiarello2021MMS,dimarco2022JSP,tosin2019MMS} and the quantification of the uncertainty in traffic data~\cite{herty2021SEMA-SIMAI,tosin2021MCMF}.

Combining classical methodologies of the kinetic theory with more modern ones developed in the above-cited papers, we will formulate optimal speed and Follow-the-Leader (FTL) non-local particle dynamics at the mesoscopic level, thereby obtaining non-local Povzner-type collisional kinetic equations of traffic. Passing then to the \textit{hydrodynamic limit}, possibly under suitable approximations of the non-locality of the interactions, we will recover explicit closure relationships at the macroscopic scale, which will result in self-consistent hydrodynamic models accounting for the non-locality in various forms. In particular, we will show that optimal speed dynamics give rise to a first order macroscopic model with a non-local flux slightly different from the ones typically postulated in heuristic constructions; and that non-local FTL dynamics have the ARZ model, or possible generalisations of it, as ``universal'' macroscopic counterpart, at least when the non-locality of the interactions is small enough.

In more detail, the paper is organised as follows. In Section~\ref{sect:optimal.speed}, we show that a class of first order non-local traffic models emerges as the hydrodynamic limit of optimal speed particle dynamics. We also compare the obtained model with other non-local first order models proposed in the literature. In Section~\ref{sect:FTL}, we show instead that the ARZ model arises as the hydrodynamic limit of FTL particle dynamics for \textit{arbitrary} non-local interaction kernels with sufficiently small support. This result generalises the one obtained in~\cite{chiarello2021IJNM}, which was based specifically on an Enskog-type kinetic description. In Section~\ref{sect:gen_FTL}, we further generalise the result of Section~\ref{sect:FTL} to a wider class of non-local FTL particle dynamics characterised by quite arbitrary interaction functions expressed in terms of the \textit{space headway} between the vehicles instead of their speed. Specifically, we show that an ARZ-like macroscopic model provides again a ``universal'' macroscopic description in suitable parameter regimes including the smallness of the support of the non-local interaction kernel. In Section~\ref{sect:numerics}, we extensively investigate and compare the solutions produced by the particle models and by the hydrodynamic models by means of numerical simulations. In Section~\ref{sect:conclusions}, we finally outline some conclusions.

\section{Optimal speed dynamics}
\label{sect:optimal.speed}
We consider a sufficiently large ensemble of indistinguishable vehicles, each of which is identified by the dimensionless position $X_t\in\R$ and dimensionless speed $V_t\in [0,\,1]$ at time $t>0$. We assume the following discrete-in-time dynamical model:
\begin{equation}
	X_{t+\Delta{t}}=X_t+V_t\Delta{t}, \qquad
		V_{t+\Delta{t}}=V_t+a\Theta\Bigl(\cV(\rho(X^\ast_t,t))-V_t\Bigr),
	\label{eq:OV_particle.1}
\end{equation}
where $\Delta{t}>0$ is a (small) time step and $a>0$ is a parameter. Moreover, $\Theta\in\{0,\,1\}$ is a binary random variable describing whether during the time step $\Delta{t}$ a randomly chosen vehicle with microscopic state $(X_t,V_t)$ updates ($\Theta=1$) or not ($\Theta=0$) its speed by relaxing it towards the \textit{optimal speed} determined by a prescribed function $\cV$. The latter depends on the traffic density $\rho$ in a point $X^\ast_t$ representing the position of another randomly picked vehicle which the previous vehicle possibly interacts with. The interaction is mediated by an \textit{interaction kernel} $B=B(X^\ast_t-X_t)$ fixing the rate at which the two vehicles with relative position $X^\ast_t-X_t$ may interact. We express this process by letting
\begin{equation}
	\Theta\sim\operatorname{Bernoulli}(B(X^\ast_t-X_t)\Delta{t}),
	\label{eq:Theta.1}
\end{equation}
so that the probability for a speed update to happen is $B(X^\ast_t-X_t)\Delta{t}$. Notice that we need $0\leq B(X^\ast_t-X_t)\Delta{t}\leq 1$ for consistency and this requires some assumptions.
\begin{assumption}[Interaction kernel] \label{ass:B}
We assume that $B$ is non-negative and compactly supported in the interval $[0,\,\eta]$ with $0<\eta<+\infty$. We also assume that $B$ is bounded.
\end{assumption}

\begin{remark}
Assumption~\ref{ass:B} implies in particular that the interaction kernel $B$ is \textit{forward-looking}. Indeed, $B(y)=0$ whenever $y<0$, thus $B(X^\ast_t-X_t)=0$ whenever $X^\ast_t<X_t$, i.e. if vehicle $X^\ast_t$ is behind vehicle $X_t$. This is consistent with the idea that interactions among vehicles are essentially \textit{anisotropic} and mainly addressed to vehicles in front. Assumption~\ref{ass:B} mimics then the behaviour of drivers who look ahead and adapt the speed of their vehicles to that of vehicles in front of them.
\end{remark}

Owing to Assumption~\ref{ass:B}, we may fix $\Delta{t}\leq\frac{1}{\sup_{y\in [0,\,\eta]}B(y)}$ in order for the law~\eqref{eq:Theta.1} of the random variable $\Theta$ to be well-defined. We point out that this is actually not a limitation, as in a moment we will consider the continuous-time limit $\Delta{t}\to 0^+$.

We also set some assumptions on the function $\cV$:
\begin{assumption}[Optimal speed] \label{ass:V}
We assume that $\cV=\cV(\rho)$ is bounded between $0$ and $1$ for all $\rho\geq 0$.
\end{assumption}

\begin{remark}
Assumption~\ref{ass:V} defines the minimal feature of $\cV$ necessary for the subsequent developments, specifically for the consistency of model~\eqref{eq:OV_particle.1}, see below. However, from the modelling point of view other characteristics may be desirable, although in our case not strictly necessary for technical purposes. Among them, we recall in particular the fact that $\cV$ be a decreasing function of $\rho$, so that the optimal speed diminishes as the traffic gets more and more congested.
\end{remark}

From Assumption~\ref{ass:V} we obtain that $0\leq a\leq 1$ is a necessary and sufficient condition for the physical consistency of the particle model~\eqref{eq:OV_particle.1}. By ``physical consistency'' we mean, in particular, that the post-interaction speed belongs to the dimensionless interval $[0,\,1]$ for any pre-interaction speed in the same interval. Writing $V_{t+\Delta{t}}=(1-a\Theta)V_t+a\Theta\cV(\rho(X^\ast_t,t))$ we recognise indeed that $V_{t+\Delta{t}}$ is a convex combination of $V_t,\,\cV(\rho(X^\ast_t,t))\in [0,\,1]$.

For completeness, we mention that the other vehicle participating in the interaction is assumed to keep its speed unchanged, hence
\begin{equation}
	X^\ast_{t+\Delta{t}}=X^\ast_t+V^\ast_t\Delta{t}, \qquad V^\ast_{t+\Delta{t}}=V^\ast_t.
	\label{eq:OV_particle.2}
\end{equation}

\subsection{Kinetic description}
\label{sect:kinetic}
To address the \textit{aggregate} trends emerging from the dynamics~\eqref{eq:OV_particle.1}-\eqref{eq:OV_particle.2} we reformulate the particle model along the lines of statistical mechanics and kinetic theory.

Let $f=f(x,v,t):\R\times [0,\,1]\times (0,\,+\infty)\to\R_+$ be the kinetic distribution function of the microscopic state $(x,v)$ of the vehicles at time $t$. Hence, $f(x,v,t)\,dx\,dv$ gives the probability that at time $t>0$ a vehicle has a position comprised between $x$ and $x+dx$ and a speed comprised between $v$ and $v+dv$. Averaging~\eqref{eq:OV_particle.1}-\eqref{eq:OV_particle.2} and taking the continuous-time limit $\Delta{t}\to 0^+$, by standard arguments (see e.g.,~\cite{fraia2020RUMI,pareschi2013BOOK}) we formally obtain that $f$ satisfies the equation
\begin{equation}
	\partial_tf+v\partial_xf=Q(f,f),
	\label{eq:kinetic.strong}
\end{equation}
where
\begin{equation}
	Q(f,f)(x,v,t):=\frac{1}{2}\int_{\R}\int_0^1B(x_\ast-x)\left(\frac{1}{J}f(x,\pr{v},t)f(x_\ast,\pr{v_\ast},t)-f(x,v,t)f(x_\ast,v_\ast,t)\right)dv_\ast\,dx_\ast
	\label{eq:Q}
\end{equation}
is the \textit{collisional operator}\footnote{``Collisional'' is a legacy from the jargon of classical kinetic theory of gases. In this context, ``collisions'' have to be meant in the abstract as ``interactions''.}. Here, $\pr{v}$, $\pr{v_\ast}$ denote the pre-interaction speeds generating the post-interaction speeds $v$, $v_\ast$ when an interaction takes place ($\Theta=1$) in the dynamics~\eqref{eq:OV_particle.1}-\eqref{eq:OV_particle.2}. Specifically,
$$ \pr{v}=v-\frac{a}{1-a}(\cV(\rho(x_\ast,t))-v), \qquad \pr{v}_\ast=v_\ast, $$
which is known as the \textit{inverse} interaction; whereas $J=1-a$ is the modulus of the Jacobian determinant of the \textit{direct} interaction, namely the transformation
\begin{equation}
	v'=v+a(\cV(\rho(x_\ast,t))-v), \qquad v'_\ast=v_\ast
	\label{eq:int.OV}
\end{equation}
from pre- to post-interaction speeds\footnote{Note the change of notation with respect to the inverse interaction, with $v$, $v_\ast$ denoting here the pre-interaction speeds and $v'$, $v'_\ast$ the post interaction speeds.}. In order for the interaction not to be singular we require $a<1$.

Due to the non-locality in space featured by $Q$,~\eqref{eq:kinetic.strong} is a \textit{Povzner-type} kinetic equation, cf.~\cite{fornasier2011PHYSD,povzner1962AMSTS}, which formally reduces to a \textit{Boltzmann-type} equation when $B$ tends to the Dirac delta centred in $x$, cf.~\cite{lachowicz1990ARMA}.

The traffic density $\rho$ appearing in~\eqref{eq:OV_particle.1} is expressed in terms of the distribution function $f$ as its zeroth-order $v$-moment:
$$ \rho(x,t):=\int_0^1f(x,v,t)\,dv. $$

\subsection{Hydrodynamic limit}
The statistical description provided by the kinetic equation~\eqref{eq:kinetic.strong} is the basis to upscale the particle dynamics~\eqref{eq:OV_particle.1}-\eqref{eq:OV_particle.2} to the macroscopic level. To this purpose, we introduce a small scale parameter $0<\eps\ll 1$, which in this context plays the role of the \textit{Knudsen number} of the classical kinetic theory, and we perform the following hyperbolic scaling of time and space:
\begin{equation}
	t\to\frac{t}{\eps}, \qquad x\to\frac{x}{\eps},
	\label{eq:hyp_scal}
\end{equation}
whence $\partial_t\to\eps\partial_t$ and $\partial_x\to\eps\partial_x$. Owing to this,~\eqref{eq:kinetic.strong} takes the form
\begin{equation}
	\partial_tf^\eps+v\partial_xf^\eps=\frac{1}{\eps}Q(f^\eps,f^\eps),
	\label{eq:kinetic.strong.eps-1st.ord}
\end{equation}
$f^\eps$ denoting now the kinetic distribution function parametrised by $\eps$.

Let $\varphi:[0,\,1]\to\R$ be an arbitrarily chosen \textit{observable quantity} (test function). From the expression~\eqref{eq:Q} of $Q$ we compute
\begin{align}
	\begin{aligned}[b]
		\int_0^1\varphi(v)&Q(f^\eps,f^\eps)(x,v,t)\,dv \\
		&=\frac{1}{2}\int_{\R}\int_0^1\int_0^1B(x_\ast-x)(\varphi(v')-\varphi(v))f^\eps(x,v,t)f^\eps(x_\ast,v_\ast,t)\,dv\,dv_\ast\,dx_\ast,
	\end{aligned}
	\label{eq:Q.weak}
\end{align}
where $v'$ is given by~\eqref{eq:int.OV}. Choosing $\varphi\equiv 1$ we obtain in particular
\begin{equation}
	\int_0^1Q(f^\eps,f^\eps)(x,v,t)\,dv=0, \qquad \forall\,\eps>0,
	\label{eq:int.Q}
\end{equation}
therefore $\varphi\equiv 1$ is a \textit{collisional invariant}. Conversely, for $\varphi(v)=v$ we obtain
\begin{equation}
	\int_0^1vQ(f^\eps,f^\eps)(x,v,t)\,dv=\frac{a}{2}\rho^\eps(x,t)\int_{\R}B(x_\ast-x)\rho^\eps(x_\ast,t)(\cV(\rho^\eps(x_\ast,t))-u^\eps(x,t))\,dx_\ast,
	\label{eq:int.vQ}
\end{equation}
where $\rho^\eps$ is the traffic density associated with the distribution function $f^\eps$ and
$$ u^\eps(x,t):=\frac{1}{\rho^\eps(x,t)}\int_0^1vf^\eps(x,v,t)\,dv $$
is its \textit{mean speed}. Thus, $\varphi(v)=v$ is in general not a collisional invariant.

Integrating~\eqref{eq:kinetic.strong.eps-1st.ord} with respect to $v$ and taking~\eqref{eq:int.Q} into account gives the conservation law
\begin{equation}
	\textcolor{red}{\partial_t}\int_0^1f^\eps(x,v,t)\,dv+\partial_x\int_0^1vf^\eps(x,v,t)\,dv=0.
	\label{eq:feps-cons.law}
\end{equation}
At the same time, passing to the \textit{hydrodynamic limit} $\eps\to 0^+$ in~\eqref{eq:kinetic.strong.eps-1st.ord} under the formal assumption that the left-hand side of the equation remains bounded implies that the limit distribution $f^0$ solves
\begin{equation}
	Q(f^0,f^0)=0
	\label{eq:local.Maxwellian}
\end{equation}
with, according to~\eqref{eq:int.vQ},
$$ u^0(x,t)=\frac{\displaystyle\int_{\R}B(x_\ast-x)\rho^0(x_\ast,t)\cV(\rho^0(x_\ast,t))\,dx_\ast}{\displaystyle\int_{\R}B(x_\ast-x)\rho^0(x_\ast,t)\,dx_\ast}. $$
In conclusion, passing formally to the hydrodynamic limit in~\eqref{eq:feps-cons.law} we discover that the traffic density $\rho^0$, which we may simply rename $\rho$ for brevity, satisfies
\begin{equation}
	\partial_t\rho+\partial_x\left(\rho\frac{\tilde{B}\ast(\rho\cV(\rho))}{\tilde{B}\ast\rho}\right)=0,
	\label{eq:1st_order.non-local}
\end{equation}
where $\tilde{B}(y):=B(-y)$ and $\ast$ denotes convolution. This is a conservation law with \textit{non-local flux} providing the hydrodynamic counterpart of the stochastic particle model~\eqref{eq:OV_particle.1}-\eqref{eq:OV_particle.2}.

\begin{remark}
It is interesting to compare~\eqref{eq:1st_order.non-local} with other non-local first order macroscopic traffic models proposed in the literature (cf.~\cite{chiarello2021CHAPTER} for a thorough overview). They are all based on the continuity equation for the traffic density $\rho$ closed with different expressions of the flux. For instance, in~\cite{blandin2016NM,chiarello2018M2AN} the following equation is proposed:
\begin{equation}
	\partial_t\rho+\partial_x\left(g(\rho)\cV\left(\int_x^{x+\eta}B(x_\ast-x)\rho(x_\ast,t)\,dx_\ast\right)\right)=0,
	\label{eq:M2}
\end{equation}
where $g$ is some non-negative function while $\cV$, $B$ are non-decreasing functions and, in particular, $B$ is a kernel modelling the behaviour of drivers adapting their speed to the density of vehicles in front of them. Instead, in~\cite{friedrich2018NHM} the following variation of the previous model is presented:
\begin{equation}
	\partial_t\rho+\partial_x\left(g(\rho)\int_x^{x+\eta}B(x_\ast-x)\cV(\rho(x_\ast,t))\,dx_\ast\right)=0,
	\label{eq:M3}
\end{equation}
where now $\cV$ is some assigned density-dependent speed. Ideally, we may say that in these models the mean traffic speed is assumed to result from:
\begin{enumerate*}[label=(\roman*)]
\item the evaluation of the average traffic density ahead in the first case;
\item the evaluation of the average traffic speed ahead in the second case.
\end{enumerate*}

Although reasonable, these models are postulated heuristically. Our derivation of~\eqref{eq:1st_order.non-local} indicates instead that a non-local first order macroscopic model consistent with simple, yet meaningful, microscopic first principles has a mean speed in the form of a spatial average of the macroscopic flux $\rho\cV(\rho)$ normalised by the corresponding spatial average of the macroscopic density $\rho$.

Finally, it is worth mentioning that in~\cite{sopasakis2006SIAP} a non-local first order macroscopic traffic model is derived from a microscopic cellular automaton implementing an asymmetric size exclusion process, which reproduces the front-rear anisotropy of vehicle interactions, and a look-ahead strategy, which accounts for non-local vehicle interactions. The resulting model reads:
$$ \partial_t\rho+\partial_x\left(\rho(1-\rho)\frac{1}{\tau}e^{-\int_x^{+\infty}B(x_\ast-x)\rho(x_\ast,t)\,dx_\ast}\right)=0, $$
where $\tau>0$ is a relaxation time. Here, the mean traffic speed is given by the classical linear diagram $1-\rho$ modulated by a non-local exponential term, which accounts for the density distribution ahead.

In Figure~\ref{fig:model_comparison}, we compare the traffic densities obtained numerically with our model~\eqref{eq:1st_order.non-local} (Model 1), model~\eqref{eq:M2} (Model 2) and model~\eqref{eq:M3} (Model 3). We reproduce the same numerical experiment proposed in \cite[Figure 6]{friedrich2018NHM} with constant interaction kernel and $\cV(\rho)=1-\rho^5$. From these numerical results we notice that our model produces less oscillations than the other models in correspondence of the rightmost discontinuity of the initial datum, namely the one which would evolve as a rarefaction wave in the local LWR model. This is consistent with the fact that, as previously mentioned, our model performs more averages of the hydrodynamic parameters compared to the other models, which may give rise to less oscillating density profiles. On the contrary, we observe that our model produces more oscillations than the other models in correspondence of the leftmost discontinuity of the initial datum, namely the one which would evolve as a shock wave in the local LWR model.

\begin{figure}[!t]
\centering
\includegraphics[width=0.4\textwidth]{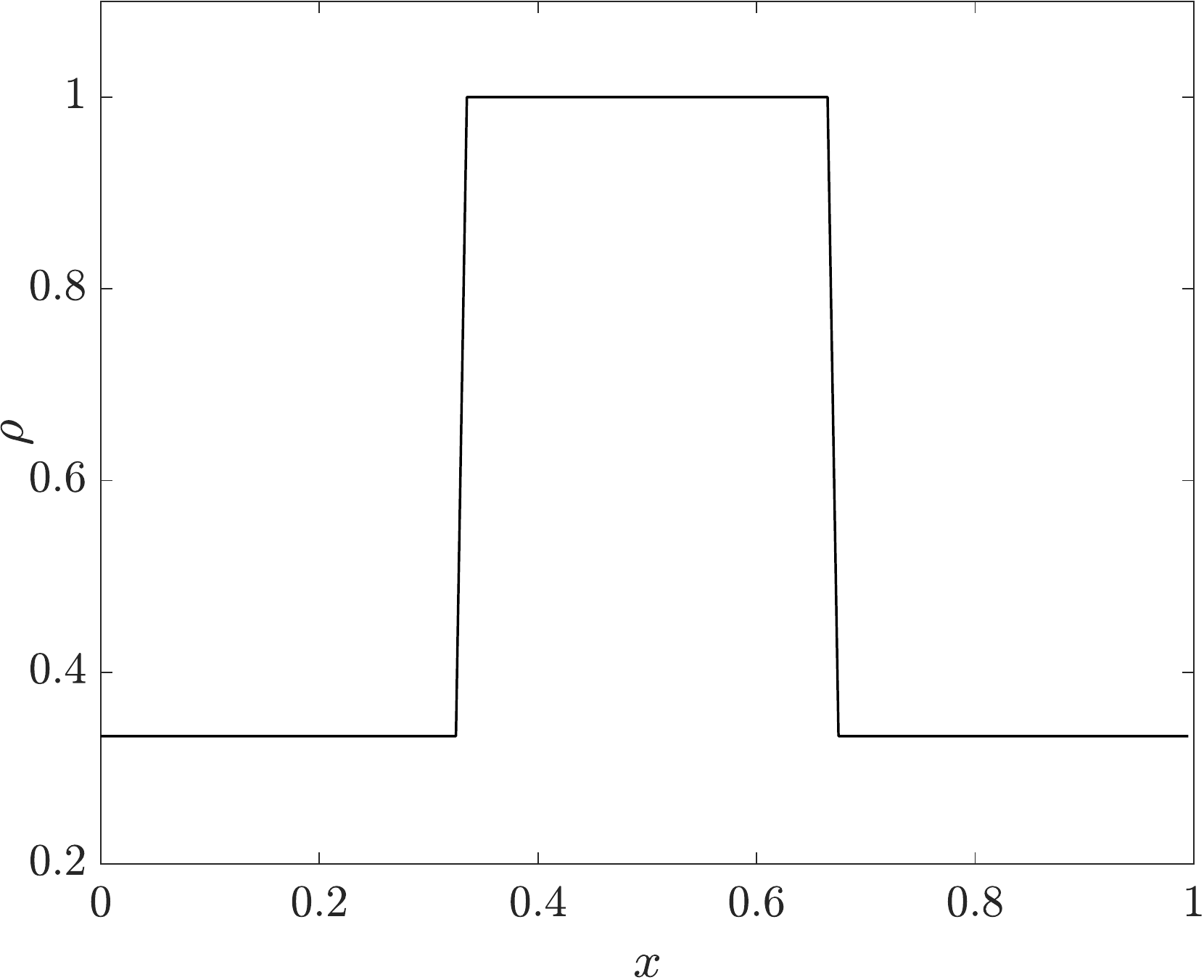}
\includegraphics[width=0.41\textwidth]{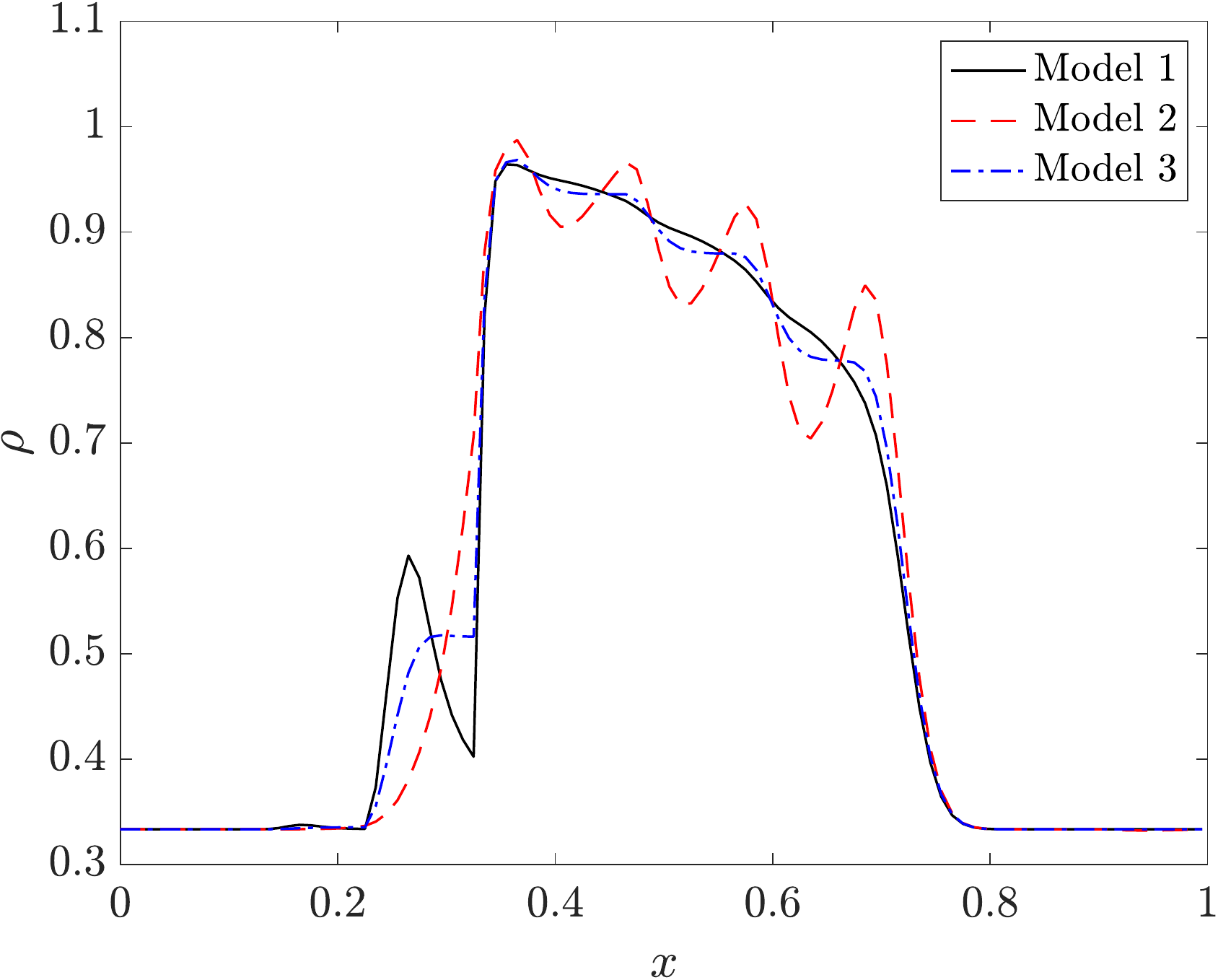} 
\caption{Numerical comparison of non-local first order macroscopic models~\eqref{eq:1st_order.non-local} (Model 1),~\eqref{eq:M2} (Model 2),~\eqref{eq:M3} (Model 3). Left: initial condition. Right: density $\rho$ at time $t=0.05$ obtained with $\cV(\rho)=1-\rho^5$, $g(\rho)=\rho$, $B(y)=\frac{1}{\eta}$ (constant interaction kernel) with $\eta=0.1$ and spatial mesh size $\Delta{x}=10^{-3}$. See~\cite[Figure 6]{friedrich2018NHM}}
\label{fig:model_comparison}
\end{figure}
\end{remark}

\section{Follow-the-Leader dynamics}
\label{sect:FTL}
In this section we consider instead \textit{Follow-the-Leader} (FTL) interactions among the vehicles, meaning that at time $t>0$ a vehicle in $X_t$ updates its speed $V_t$ on the basis of the speed $V^\ast_t$ of another vehicle in $X^\ast_t$. Unlike~\eqref{eq:OV_particle.1}, now the speed $V^\ast_t$ plays a direct role in determining the new speed $V_{t+\Delta{t}}$. Although the interacting vehicle $(X^\ast_t,V^\ast_t)$ is chosen randomly in the traffic stream, it is usually though of as a leading vehicle of vehicle $(X_t,V_t)$.

The particle model~\eqref{eq:OV_particle.1} modifies as
\begin{equation}
	X_{t+\Delta{t}}=X_t+V_t\Delta{t}, \qquad
		V_{t+\Delta{t}}=V_t+\lambda\Theta(V^\ast_t-V_t),
	\label{eq:FTL_particle.1}
\end{equation}
where $\lambda>0$ is a constant coefficient representing the sensitivity of the drivers, cf.~\cite{gazis1961OR}. Again,
$$ \Theta\sim\operatorname{Bernoulli}(B(X^\ast_t-X_t)\Delta{t}) $$
with the interaction kernel $B$ fulfilling Assumption~\ref{ass:B}. Rules~\eqref{eq:FTL_particle.1} are complemented also in this case with
\begin{equation}
	X^\ast_{t+\Delta{t}}=X^\ast_t+V^\ast_t\Delta{t}, \qquad V^\ast_{t+\Delta{t}}=V^\ast_t,
	\label{eq:FTL_particle.2}
\end{equation}
sticking to the idea that in an interaction the leading vehicle does not change speed.

Condition $0\leq\lambda\leq 1$ is necessary and sufficient to guarantee the physical consistency of~\eqref{eq:FTL_particle.1}, specifically the fact that $V_{t+\Delta{t}}\in [0,\,1]$ for all $V_t,\,V^\ast_t\in [0,\,1]$. Indeed, $V_{t+\Delta{t}}$ is a convex combination of $V_t$, $V^\ast_t$ through the coefficient $\lambda\Theta$.

\subsection{Kinetic description}
The kinetic description is provided by~\eqref{eq:kinetic.strong},~\eqref{eq:Q}, the inverse interaction being now
$$ \pr{v}=v-\frac{\lambda}{1-\lambda}(v_\ast-v), \qquad \pr{v}_\ast=v_\ast $$
and the direct interaction
\begin{equation}
	v'=v+\lambda(v_\ast-v), \qquad v'_\ast=v_\ast
	\label{eq:int.FTL}
\end{equation}
with Jacobian determinant $J=1-\lambda$. In order for the interaction not to be singular we require $\lambda<1$.

For purposes which will be clear in a moment, if the spatial non-locality of the interactions is sufficiently small, namely if
\begin{equation}
	\eta=\abs{\supp{B}}\ll 1,
	\label{eq:eta.small}
\end{equation}
cf. Assumption~\ref{ass:B}, we may introduce an approximation of the collisional operator $Q$, which will be useful in the sequel. Consider the first order truncation of the Taylor expansion of $f$ about $x$:
\begin{equation}
	f(x_\ast,v_\ast,t)\approx f(x,v_\ast,t)+\partial_xf(x,v_\ast,t)(x_\ast-x),
	\label{eq:f.approx}
\end{equation}
which is justified by the fact that if $x_\ast-x \in\supp{B}$ then, owing to~\eqref{eq:eta.small}, $x,\,x_\ast$ are close to each other. Consequently, from~\eqref{eq:Q} we obtain:
\begin{align}
	\begin{aligned}[b]
		Q(f,f)(x,v,t) &\approx \frac{\cB_0}{2}\int_0^1\left(\frac{1}{J}f(x,\pr{v},t)f(x,\pr{v_\ast},t)-f(x,v,t)f(x,v_\ast,t)\right)dv_\ast \\
		&\phantom{=} +\frac{\cB_1}{2}\int_0^1\left(\frac{1}{J}f(x,\pr{v},t)\partial_xf(x,\pr{v_\ast},t)-f(x,v,t)\partial_xf(x,v_\ast,t)\right)dv_\ast \\
		&=: \cB_0Q_\B(f,f)(x,v,t)+\cB_1Q_\B(f,\partial_xf)(x,v,t),
	\end{aligned}
	\label{eq:Q.approx}
\end{align}
where
\begin{equation}
	\cB_0:=\int_{\R}B(y)\,dy, \qquad \cB_1:=\int_{\R}yB(y)\,dy.
	\label{eq:B0.B1}
\end{equation}
Notice that $Q_\B(f,f)$ is a classical Boltzmann-type collisional operator in which the distribution functions of the interacting vehicles are computed in the same space position $x$. The local correction $Q_\B(f,\partial_xf)$ keeps instead track of the (small) non-locality of the interactions. In weak form, for an arbitrary observable quantity $\varphi=\varphi(v)$, we have (cf.~\eqref{eq:Q.weak}):
\begin{equation}
	\int_0^1\varphi(v)Q_\B(f,g)(x,v,t)\,dv=\frac{1}{2}\int_0^1\int_0^1(\varphi(v')-\varphi(v))f(x,v,t)g(x,v_\ast,t)\,dv\,dv_\ast
	\label{eq:Q.approx-weak}
\end{equation}
with $g=f$ and $g=\partial_xf$, respectively.

\subsection{Hydrodynamic limit}
By the hyperbolic scaling~\eqref{eq:hyp_scal} of space and time we obtain the scaled kinetic equation~\eqref{eq:kinetic.strong.eps-1st.ord}. Next, from~\eqref{eq:Q.weak} we may identify the collisional invariants, using now the interaction rules~\eqref{eq:int.FTL}. In particular, we get again that $\varphi\equiv 1$ is a collisional invariant, while for $\varphi(v)=v$ we obtain
$$ \int_0^1vQ(f^\eps,f^\eps)(x,v,t)\,dv=\frac{\lambda}{2}\rho^\eps(x,t)\int_{\R}B(x_\ast-x)\rho^\eps(x_\ast,t)(u^\eps(x_\ast,t)-u^\eps(x,t))\,dx_\ast. $$
We notice that, even for $\eps\to 0^+$ when~\eqref{eq:local.Maxwellian} holds, from this relationship it is not immediate to extract explicit analytical information on the limit mean speed $u^0$. Thus it is difficult to proceed with the classical hydrodynamic limit, which requires to identify precisely the collisional invariants and the local equilibrium values of the non-conserved quantities. To circumvent this difficulty, we assume~\eqref{eq:eta.small} and look for the best local hydrodynamic approximation of the non-local FTL particle dynamics.

\subsubsection{Approximate hydrodynamics}
\label{sect:approx.hydro}
Under assumption~\eqref{eq:eta.small} we may rely on the approximation~\eqref{eq:Q.approx} of the collisional operator $Q$. The hyperbolic scaling~\eqref{eq:hyp_scal} causes the kinetic equation~\eqref{eq:kinetic.strong} to take the form
\begin{equation}
	\partial_tf^\eps+v\partial_xf^\eps=\frac{\cB_0}{\eps}Q_\B(f^\eps,f^\eps)+\cB_1Q_\B(f^\eps,\partial_xf^\eps),
	\label{eq:kinetic.strong.eps-2nd.ord}
\end{equation}
considering that $Q_\B(f^\eps,\eps\partial_xf^\eps)=\eps Q_\B(f^\eps,\partial_xf^\eps)$ from the bi-linearity of $Q_\B$. Recalling~\eqref{eq:int.FTL} and~\eqref{eq:Q.approx-weak}, we easily see that
$$ \int_0^1Q_\B(f^\eps,f^\eps)(x,v,t)\,dv=\int_0^1vQ_\B(f^\eps,f^\eps)(x,v,t)\,dv=0, \qquad \forall\,\eps>0, $$
hence $\varphi\equiv 1$ and $\varphi(v)=v$ are collisional invariants. Conversely,
\begin{equation}
	\int_0^1v^2Q_\B(f^\eps,f^\eps)(x,v,t)\,dv=\lambda(1-\lambda){(\rho^\eps)}^2(x,t)\Bigl({(u^\eps)}^2(x,t)-E^\eps(x,t)\Bigr),
	\label{eq:int.v^2Q}
\end{equation}
where we have denoted by
$$ E^\eps=E^\eps(x,t):=\frac{1}{\rho^\eps(x,t)}\int_0^1v^2f^\eps(x,v,t)\,dv $$
the \textit{energy} of the distribution function $f^\eps$. Parallelly, from~\eqref{eq:Q.approx-weak} we observe that
\begin{align*}
	& \int_0^1Q_\B(f^\eps,\partial_xf^\eps)(x,v,t)\,dv=0 \\
	& \int_0^1vQ_\B(f^\eps,\partial_xf^\eps)(x,v,t)\,dv=\frac{\lambda}{2}\left(\int_0^1f^\eps(x,v,t)\,dv\cdot\partial_x\int_0^1v_\ast f^\eps(x,v_\ast,t)\,dv_\ast\right. \\
	& \phantom{\int_0^1vQ_\B(f^\eps,\partial_xf^\eps)(x,v,t)\,dv=} \left.-\partial_x\int_0^1f^\eps(x,v_\ast,t)\,dv_\ast\cdot\int_0^1vf^\eps(x,v,t)\,dv\right), \qquad \forall\,\eps>0.
\end{align*}
Therefore, multiplying~\eqref{eq:kinetic.strong.eps-2nd.ord} by the two collisional invariants above and integrating with respect to $v$ yields, for every $\eps>0$, the following system  of balance laws:
\begin{align}
	\resizebox{.9\hsize}{!}{$
	\left\{
	\begin{aligned}[c]
		& \partial_t\int_0^1f^\eps(x,v,t)\,dv+\partial_x\int_0^1vf^\eps(x,v,t)\,dv=0 \\
		& \partial_t\int_0^1vf^\eps(x,v,t)\,dv+\partial_x\int_0^1v^2f^\eps(x,v,t)\,dv=\frac{\lambda\cB_1}{2}\left(\int_0^1f^\eps(x,v,t)\,dv\cdot\partial_x\int_0^1v_\ast f^\eps(x,v_\ast,t)\,dv_\ast\right. \\
		& \phantom{\partial_t\int_0^1vf^\eps(x,v,t)\,dv+\partial_x\int_0^1v^2f^\eps(x,v,t)\,dv=} \left.-\int_0^1vf^\eps(x,v,t)\,dv\cdot\partial_x\int_0^1f^\eps(x,v_\ast,t)\,dv_\ast\right).
	\end{aligned}
	\right.
	$}
	\label{eq:feps-balance.laws}
\end{align}
Taking now to the limit $\eps\to 0^+$ in~\eqref{eq:kinetic.strong.eps-2nd.ord} and assuming formally that the left-hand side of the equation, as well as the second term at the right-hand side, remain bounded we get that the limit distribution $f^0$ solves
$$ Q_\B(f^0,f^0)=0 $$
with, owing to~\eqref{eq:int.v^2Q}, $E^0(x,t)={(u^0)}^2(x,t)$ at least for $0<\lambda<1$ (notice that $\lambda<1$ by assumption while $\lambda>0$ is necessary for meaningfulness of the model, as $\lambda=0$ would imply no interactions among the vehicles). Consequently, passing formally to the limit in~\eqref{eq:feps-balance.laws} we obtain that the traffic density and mean speed $\rho^0$, $u^0$, which we rename simply $\rho$, $u$ for convenience, satisfy
\begin{equation*}
	\begin{cases}
		\partial_t\rho+\partial_x(\rho u)=0 \\
		\partial_t(\rho u)+\partial_x(\rho u^2)=\dfrac{\lambda\cB_1}{2}\rho^2\partial_xu.
	\end{cases}
\end{equation*}
Using the first equation, we rewrite the second equation as
$$ \partial_tu+\left(u-\rho\frac{\lambda\cB_1}{2}\right)\partial_xu=0 $$
so that, on the whole, the hydrodynamic description of the particle model~\eqref{eq:FTL_particle.1}-\eqref{eq:FTL_particle.2} under the assumption~\eqref{eq:eta.small} of small non-locality of the interactions is provided by the system
\begin{equation}
	\begin{cases}
		\partial_t\rho+\partial_x(\rho u)=0 \\[2mm]
		\partial_tu+\left(u-\rho\dfrac{\lambda\cB_1}{2}\right)\partial_xu=0,
	\end{cases}
	\label{eq:ARZ}
\end{equation}
i.e. the celebrated \textit{Aw-Rascle-Zhang} (ARZ) \textit{traffic model}~\cite{aw2000SIAP,zhang2002TRB} with (pseudo-) pressure $p=p(\rho)$ satisfying
$$ p'(\rho)=\frac{\lambda\cB_1}{2}. $$
Hence, $p(\rho):=\frac{\lambda\cB_1}{2}\rho$ up to an arbitrary additive constant. It is immediate to check that~\eqref{eq:ARZ} is a hyperbolic model with eigenvalues $u$ and $u-\rho\frac{\lambda\cB_1}{2}$. From Assumption~\ref{ass:B}, in particular the fact that $\supp{B}\subseteq\R_+$, and~\eqref{eq:B0.B1} we deduce $\cB_1>0$, hence no eigenvalue of~\eqref{eq:ARZ} is greater than $u$. Consequently, model~\eqref{eq:ARZ} has the property that small disturbances in traffic cannot propagate faster than the flow of vehicles, which reproduces correctly the front-rear anisotropy of vehicle interactions. Fulfilling such a property was the main motivation for the introduction of the ARZ model~\cite{aw2000SIAP} as a cure for some physical inconsistencies of previous second order macroscopic traffic models~\cite{daganzo1995TR}.

Summarising, we have shown that:
\begin{quotation}
\it The hydrodynamic limit of \emph{any} non-local FTL particle model of the form~\eqref{eq:FTL_particle.1}-\eqref{eq:FTL_particle.2} \emph{with sufficiently small non-locality} is the ARZ model~\eqref{eq:ARZ} with a (pseudo-) pressure $p$ reminiscent of the non-local interaction kernel through the coefficient $\cB_1$.
\end{quotation}

This result generalises the one obtained in~\cite{dimarco2020JSP}, where the ARZ model is recovered as the hydrodynamic limit of an \textit{Enskog-type} kinetic description of FTL vehicle dynamics. Notice that an Enskog-type description amounts to~\eqref{eq:kinetic.strong}-\eqref{eq:Q} with the particular choice $B(y)=\delta(y-\eta)$, where $\delta$ denotes the Dirac delta distribution and $\eta>0$ is in this case the fixed distance separating the interacting vehicles.

This result shows also that the ARZ model, originally conceived as a macroscopic traffic model accounting for the anticipation ability of the drivers in a physically consistent way, cf.~\cite{aw2000SIAP,daganzo1995TR,rascle2002MCM}, is in fact more in general the prototype of any non-local FTL interaction model at least for a sufficiently small non-locality.

\section{Generalised Follow-the-Leader dynamics}
\label{sect:gen_FTL}
In this section we consider the following generalisation of the non-local FTL particle model~\eqref{eq:FTL_particle.1}-\eqref{eq:FTL_particle.2}:
\begin{equation}
	X_{t+\Delta{t}}=X_t+\cV(S_t)\Delta{t}, \qquad S_{t+\Delta{t}}=S_t+\Theta(\Psi_\lambda(S^\ast_t)-\Psi_\lambda(S_t)),
	\label{eq:sFTL_particle.1}
\end{equation}
and
\begin{equation}
	X^\ast_{t+\Delta{t}}=X^\ast_t+\cV(S^\ast_t)\Delta{t}, \qquad S^\ast_{t+\Delta{t}}=S^\ast_t,
	\label{eq:sFTL_particle.2}
\end{equation}
where now $S_t,\,S^\ast_t\geq 0$ are the \textit{space headways} of the interacting vehicles, i.e. the free space each of them has in front, $\cV:\R_+\to\ [0,\,1]$ is the speed of a vehicle given as a function of the headway and $\Psi_\lambda:\R_+\to\R_+$ is an interaction function parametrised by the sensitivity $\lambda>0$ of the drivers. Follow-the-Leader particle models that can be recast in the form~\eqref{eq:sFTL_particle.1} have been considered e.g., in~\cite{chiarello2021MMS,piccoli2020ZAMP,tosin2021SEMA-SIMAI}.

We set the following technical assumptions on $\Psi_\lambda$:
\begin{assumption}[Interaction function] \label{ass:Psi}
We assume that $\Psi_\lambda$ is non-negative and such that $\Psi_\lambda(s)\sim\lambda s$ when $\lambda\to 0^+$. In particular, we assume that there exists $\psi_\lambda:\R_+\to\R_+$, with $\psi_\lambda(s)\to 1$ for all $s\in\R_+$ when $\lambda\to 0^+$, such that
$$ \Psi_\lambda(s)=\lambda s\psi_\lambda(s). $$
Moreover, we assume
$$ \sup_{\lambda>0}\sup_{s\in\R_+}\psi_\lambda(s)<+\infty. $$
\end{assumption}

A reference family of functions $\Psi_\lambda$ complying with Assumption~\ref{ass:Psi} is
$$ \Psi_\lambda(s)=\frac{\lambda s}{{(1+\lambda s^\alpha)}^\beta} $$
with $\alpha,\,\beta>0$. In the aforementioned papers~\cite{chiarello2021MMS,piccoli2020ZAMP,tosin2021SEMA-SIMAI} the case $\alpha=\beta=1$ is considered with the interaction function written in the equivalent form $\Psi_\lambda(s)=1-\frac{1}{1+\lambda s}$.

Unlike the cases discussed in the previous sections, now it is not possible to identify a universal range of values of $\lambda$, valid for all possible choices of $\Psi_\lambda$, which ensures the physical consistency of interactions~\eqref{eq:sFTL_particle.1}, namely the fact that $S_{t+\Delta{t}}\geq 0$ for all $S_t,\,S^\ast_t\geq 0$. Therefore it is necessary to neglect explicitly possible unphysical interactions producing $S_{t+\Delta{t}}<0$. This may be done by applying a \textit{cutoff} to the interaction frequency parametrising the law of $\Theta$:
\begin{equation}
	\Theta\sim\operatorname{Bernoulli}\left(\frac{1}{\lambda}\chi(S_t+\Psi_\lambda(S^\ast_t)-\Psi_\lambda(S_t)\geq 0)B(X^\ast_t-X_t)\Delta{t}\right),
	\label{eq:Theta.3}
\end{equation}
where $\chi(\cdot)$ is the characteristic function of the event in parenthesis. $S_t+\Psi_\lambda(S^\ast_t)-\Psi_\lambda(S_t)$ is the value of the post-interaction headway if an interaction takes place. If the pair $(S_t,\,S^\ast_t)$ produces an unphysical $S_{t+\Delta{t}}<0$ then the corresponding interaction between the two vehicles is discarded, because $\chi(S_t+\Psi_\lambda(S^\ast_t)-\Psi_\lambda(S_t)\geq 0)=0$ whence $\Theta\sim\operatorname{Bernoulli}(0)$. The further coefficient $\frac{1}{\lambda}$ in~\eqref{eq:Theta.3} is used to normalise the interaction rate with respect to the sensitivity of the drivers, in such a way that different models corresponding to different $\Psi_\lambda$ are more comparable.

\subsection{Kinetic description}
The kinetic description of the particle model~\eqref{eq:sFTL_particle.1}-\eqref{eq:Theta.3} is now provided by a distribution function $f=f(x,s,t):\R\times\R_+\times (0,\,+\infty)\to\R_+$ of the microscopic state $(x,s)$ of the vehicles. By the same guidelines recalled in Section~\ref{sect:kinetic}, we obtain that the kinetic equation satisfied by $f$ is of the form~\eqref{eq:kinetic.strong} with
\begin{multline}
	Q(f,f)(x,s,t)= \\ \frac{1}{2\lambda}\int_\R\int_{\R_+}B(x_\ast-x)\left(\chi(s\geq 0)\frac{1}{J}f(x,\pr{s},t)f(x_\ast,\pr{s}_\ast,t)-\chi(s'\geq 0)f(x,s,t)f(x_\ast,s_\ast,t)\right)ds_\ast\,dx_\ast,
	\label{eq:Q.cutoff}
\end{multline}
where:
\begin{enumerate}[label=(\roman*)]
\item $\pr{s}$, $\pr{s}_\ast$ are the pre-interaction headways producing the post-interaction headways $s$, $s_\ast$ if an interaction takes place ($\Theta=1$) in the dynamics~\eqref{eq:sFTL_particle.1}-\eqref{eq:Theta.3}. The inverse interaction giving $\pr{s}$, $\pr{s}_\ast$ as functions of $s$, $s_\ast$ is
$$ \pr{s}-\Psi_\lambda(\pr{s})=s-\Psi_\lambda(s_\ast), \qquad \pr{s}_\ast=s_\ast; $$
more explicit expressions depend on the specific function $\Psi_\lambda$ used;
\item $J=\abs{1-\Psi_\lambda'(\pr{s})}$ is the modulus of the Jacobian determinant of the direct interaction, i.e.
\begin{equation}
	s'=s+\Psi_\lambda(s_\ast)-\Psi_\lambda(s), \qquad s_\ast'=s_\ast,
	\label{eq:int.sFTL}
\end{equation}
which is also used in the term $\chi(s'\geq 0)$;
\item the coefficient $\frac{1}{\lambda}$ in front of the expression of $Q$ comes from normalisation factor of the interaction rate in~\eqref{eq:Theta.3}.
\end{enumerate}

The weak form of the collisional operator $Q$ reads
\begin{align}
	\begin{aligned}[b]
		\int_{\R_+}\varphi(s)&Q(f,f)(x,s,t)\,ds \\
		&= \frac{1}{2\lambda}\int_\R\int_{\R_+}\int_{\R_+}B(x_\ast-x)\chi(s'\geq 0)(\varphi(s')-\varphi(s))f(x,s,t)f(x_\ast,s_\ast,t)\,ds\,ds_\ast\,dx_\ast
	\end{aligned}
	\label{eq:Q.cutoff-weak}
\end{align}
for every observable quantity $\varphi:\R_+\to\R$. This expression is similar to~\eqref{eq:Q} but now the full interaction kernel is $B(x_\ast-x)\chi(s'\geq 0)$. In particular, the cutoff $\chi(s'\geq 0)$ introduces a strong non-linearity in $Q$, which makes the kinetic equation less amenable to analytical investigations. To circumvent this difficulty of the theory, we consider the particle model~\eqref{eq:sFTL_particle.1}-\eqref{eq:Theta.3} and the corresponding kinetic description in the limit $\lambda\to 0^+$. The motivation for taking $\lambda$ small is that~\eqref{eq:sFTL_particle.1} and Assumption~\ref{ass:Psi} suggest that the smaller $\lambda$ the easier to meet the requirement $s'\geq 0$ because $s'\approx s$. Consequently, we may expect the cutoff in~\eqref{eq:Q.cutoff-weak} to be less and less problematic. Since, as we mentioned before, a universal maximum value of $\lambda$ cannot be fixed, the limit $\lambda\to 0^+$ is also representative of universal aggregate trends at least in the regime of a small sensitivity of the drivers.

Since $\chi(s'\geq 0)=1-\chi(s'<0)$, we may rewrite~\eqref{eq:Q.cutoff-weak} as
\begin{align*}
	\int_{\R_+}\varphi(s)&Q(f,f)(x,s,t)\,ds \\
	&= \frac{1}{2\lambda}\int_\R\int_{\R_+}\int_{\R_+}B(x_\ast-x)(\varphi(s')-\varphi(s))f(x,s,t)f(x_\ast,s_\ast,t)\,ds\,ds_\ast\,dx_\ast \\
	&\phantom{=} -\frac{1}{2\lambda}\int_\R\int_{\R_+}\int_{\R_+}B(x_\ast-x)\chi(s'<0)(\varphi(s')-\varphi(s))f(x,s,t)f(x_\ast,s_\ast,t)\,ds\,ds_\ast\,dx_\ast \\
	&=: \cQ(f,f)[\varphi](x,t)+\cR(f,f)[\varphi](x,t).
\end{align*}
Now observe from~\eqref{eq:int.sFTL} that $s-\Psi_\lambda(s)\leq s'$ and furthermore, owing to Assumption~\ref{ass:Psi}, $s-\Psi_\lambda(s)=s-\lambda s\psi_\lambda(s)\geq (1-\lambda C)s$, where $C>0$ is a constant such that $\psi_\lambda(s)\leq C$ for all $\lambda,\,s\in\R_+$. Hence $(1-\lambda C)s\leq s'$, which implies $\chi(s'<0)\leq\chi((1-\lambda C)s<0)$. Consequently,
\begin{multline*}
	\abs{\cR(f,f)[\varphi](x,t)}\leq \\
		\frac{1}{2\lambda}\int_\R\int_{\R_+}\int_{\R_+}B(x_\ast-x)\chi((1-\lambda C)s<0)\abs{\varphi(s')-\varphi(s)}f(x,s,t)f(x_\ast,s_\ast,t)\,ds\,ds_\ast\,dx_\ast
\end{multline*}
and since $\chi((1-\lambda C)s<0)=0$ for $\lambda\leq\frac{1}{C}$ we get
$$ \lim_{\lambda\to 0^+}\cR(f,f)[\varphi](x,t)=0, $$
whence
$$ \lim_{\lambda\to 0^+}\int_{\R_+}\varphi(s)Q(f,f)(x,s,t)\,ds=\lim_{\lambda\to 0^+}\cQ(f,f)[\varphi](x,t). $$
To compute the limit of $\cQ(f,f)[\varphi]$ we assume that $\varphi$ is sufficiently smooth, say $\varphi\in C^2_c(\R_+)$, and we expand $\varphi(s')-\varphi(s)$ in Taylor series up to the second order with Lagrange remainder:
\begin{align*}
	\cQ(f,f)&[\varphi](x,t) \\
	&= \frac{1}{2}\int_\R\int_{\R_+}\int_{\R_+}B(x_\ast-x)\varphi'(s)\bigl(s_\ast\psi_\lambda(s_\ast)-s\psi_\lambda(s)\bigr)f(x,s,t)f(x_\ast,s_\ast,t)\,ds\,ds_\ast\,dx_\ast \\
	&\phantom{=} +\frac{\lambda}{2}\int_\R\int_{\R_+}\int_{\R_+}B(x_\ast-x)\varphi''(\bar{s}){\bigl(s_\ast\psi_\lambda(s_\ast)-s\psi_\lambda(s)\bigr)}^2f(x,s,t)f(x_\ast,s_\ast,t)\,ds\,ds_\ast\,dx_\ast,
\end{align*}
where $\bar{s}:=\vartheta s+(1-\vartheta)s_\ast$ for some $\vartheta\in [0,\,1]$. Since, owing to Assumption~\ref{ass:Psi}, $\psi_\lambda$ is uniformly bounded with respect to $\lambda,\,s\in\R_+$, and, owing to Assumption~\ref{ass:B}, $B$ is also bounded, we may pass to the limit under the integral by dominated convergence up to assuming that $f$ has $x$-integrable first and second $s$-moments, i.e.:
$$ \int_\R\int_{\R_+}s^pf(x,s,t)\,ds\,dx<+\infty \quad \text{for } p=1,\,2. $$
In such a case we obtain
\begin{align*}
	\lim_{\lambda\to 0^+}\cQ(f,f)[\varphi](x,t) &= \frac{1}{2}\int_\R\int_{\R_+}\int_{\R_+}B(x_\ast-x)\varphi'(s)(s_\ast-s)f(x,s,t)f(x_\ast,s_\ast,t)\,ds\,ds_\ast\,dx_\ast \\
	&=: \int_{\R_+}\varphi(s)\tilde{Q}(f,f)(x,s,t)\,ds,
\end{align*}
$\tilde{Q}(f,f)$ being the operator by which we may replace $Q(f,f)$ given by~\eqref{eq:Q.cutoff} in the limit $\lambda\to 0^+$.

Under~\eqref{eq:eta.small} and~\eqref{eq:f.approx} we may further approximate
\begin{align}
	\begin{aligned}[b]
		\int_{\R_+}\varphi(s)\tilde{Q}(f,f)(x,s,t)\,ds &\approx \frac{\cB_0}{2}\int_{\R_+}\int_{\R_+}\varphi'(s)(s_\ast-s)f(x,s,t)f(x,s_\ast,t)\,ds\,ds_\ast \\
		&\phantom{\approx} +\frac{\cB_1}{2}\int_{\R_+}\int_{\R_+}\varphi'(s)(s_\ast-s)f(x,s,t)\partial_xf(x,s_\ast,t)\,ds\,ds_\ast \\
		&=: \cB_0\int_{\R_+}\varphi(s)\tilde{Q}_\B(f,f)(x,s,t)\,ds+\cB_1\int_{\R_+}\varphi(s)\tilde{Q}_\B(f,\partial_xf)(x,s,t)\,ds,
	\end{aligned}
	\label{eq:Q_tilde.weak}
\end{align}
where $\tilde{Q}_\B(f,f)$ expresses local interactions while $\tilde{Q}_\B(f,\partial_xf)$ is a correction approximating the small non-locality by means of a space derivative.

The kinetic equation~\eqref{eq:kinetic.strong} takes finally the form
\begin{equation}
	\partial_tf+\cV(s)\partial_xf=\cB_0\tilde{Q}_\B(f,f)+\cB_1\tilde{Q}_\B(f,\partial_xf),
	\label{eq:Vlasov.strong}
\end{equation}
considering that, according to~\eqref{eq:sFTL_particle.1},~\eqref{eq:sFTL_particle.2}, the transport speed of the vehicles is $\cV=\cV(s)$.

\subsection{Hydrodynamic limit}
The hyperbolic scaling~\eqref{eq:hyp_scal} applied to~\eqref{eq:Vlasov.strong} produces the equation
\begin{equation}
	\partial_tf^\eps+\cV(s)\partial_xf^\eps=\frac{\cB_0}{\eps}\tilde{Q}_\B(f^\eps,f^\eps)+\cB_1\tilde{Q}_\B(f^\eps,\partial_xf^\eps),
	\label{eq:kinetic.strong.eps-2nd.ord-s}
\end{equation}
which in the hydrodynamic limit $\eps\to 0^+$ can be tackled similarly to Section~\ref{sect:approx.hydro}. In particular, letting $\varphi(s)=1,\,s$ in~\eqref{eq:Q_tilde.weak} we discover
$$ \int_{\R_+}\tilde{Q}_\B(f^\eps,f^\eps)(x,s,t)\,ds=\int_{\R_+}s\tilde{Q}_\B(f^\eps,f^\eps)(x,s,t)\,ds=0, \qquad \forall\,\eps>0, $$
whereas for $\varphi(s)=s^2$ we obtain
\begin{equation}
	\int_{\R_+}s^2\tilde{Q}_\B(f^\eps,f^\eps)(x,s,t)\,ds={(\rho^\eps)}^2(x,t)\Bigl({(h^\eps)}^2(x,t)-E^\eps(x,t)\Bigr)
	\label{eq:int.s^2Q}
\end{equation}
with
$$ h^\eps(x,t):=\frac{1}{\rho^\eps(x,t)}\int_{\R_+}sf^\eps(x,s,t)\,ds, \qquad	E^\eps(x,t):=\frac{1}{\rho^\eps(x,t)}\int_{\R_+}s^2f^\eps(x,s,t)\,ds $$
the \textit{mean headway} and the energy of the distribution $f^\eps$, respectively.

Multiplying~\eqref{eq:kinetic.strong.eps-2nd.ord-s} by $\varphi(s)=1,\,s$ and integrating with respect to $s$ produces, for every $\eps>0$, the following system of balance laws:
\begin{align}
	\resizebox{.9\hsize}{!}{$
	\left\{
	\begin{aligned}[c]
		& \partial_t\int_0^1f^\eps(x,s,t)\,ds+\partial_x\int_0^1\cV(s)f^\eps(x,s,t)\,ds=0 \\
		& \partial_t\int_0^1sf^\eps(x,s,t)\,ds+\partial_x\int_0^1s\cV(s)f^\eps(x,s,t)\,ds=\frac{\cB_1}{2}\left(\int_0^1f^\eps(x,s,t)\,ds\cdot\partial_x\int_0^1s_\ast f^\eps(x,s_\ast,t)\,ds_\ast\right. \\
		& \phantom{\partial_t\int_0^1sf^\eps(x,s,t)\,ds+\partial_x\int_0^1s\cV(s)f^\eps(x,s,t)\,ds=} \left.-\int_0^1sf^\eps(x,s,t)\,ds\cdot\partial_x\int_0^1f^\eps(x,s_\ast,t)\,ds_\ast\right).
	\end{aligned}
	\right.
	$}
	\label{eq:feps-balance.laws-s}
\end{align}
Taking instead the limit $\eps\to 0^+$ in~\eqref{eq:kinetic.strong.eps-2nd.ord-s}, while assuming formally that the left-hand side and the second term at the right-hand side remain bounded, yields that the limit distribution $f^0$ solves
$$ \tilde{Q}_\B(f^0,f^0)=0 $$
with, owing to~\eqref{eq:int.s^2Q}, $E^0(x,t)={(h^0)}^2(x,t)$. This implies in particular that $f^0$ has zero variance, hence that it is the \textit{monokinetic} distribution
$$ f^0(x,s,t)=\rho^0(x,t)\delta(s-h^0(x,t)). $$
Consequently, passing formally to the limit in~\eqref{eq:feps-balance.laws-s}, we discover that the traffic density and mean headway $\rho^0$, $h^0$, which we rename simply as $\rho$, $h$, satisfy
\begin{equation*}
	\begin{cases}
		\partial_t\rho+\partial_x(\rho\cV(h))=0 \\
		\partial_t(\rho h)+\partial_x(\rho h\cV(h))=\dfrac{\cB_1}{2}\rho^2\partial_xh.
	\end{cases}
\end{equation*}
Using the first equation, we can rewrite the second equation as
$$ \partial_th+\left(\cV(h)-\rho\dfrac{\cB_1}{2}\right)\partial_xh=0, $$
whereby the hydrodynamic description of the particle model~\eqref{eq:sFTL_particle.1}-\eqref{eq:Theta.3} in the limit $\lambda\to 0^+$ takes finally the form
\begin{equation}
	\begin{cases}
		\partial_t\rho+\partial_x(\rho\cV(h))=0 \\[2mm]
		\partial_th+\left(\cV(h)-\rho\dfrac{\cB_1}{2}\right)\partial_xh=0.
	\end{cases}
	\label{eq:generalised_ARZ}
\end{equation}

We notice that model~\eqref{eq:generalised_ARZ} has evident formal analogies with the ARZ model~\eqref{eq:ARZ}. In particular, it is easy to see that also this model is hyperbolic with eigenvalues given by $\cV(h)$ and $\cV(h)-\rho\frac{\cB_1}{2}$. Since now $\cV(h)$ is the flow speed and, as already noticed, $\cB_1>0$, we recover the physically meaningful property that small disturbances in traffic do not propagate faster than the flow of the vehicles.

In conclusion:
\begin{quotation}
\it In the regime of small driver sensitivity $\lambda$, the hydrodynamic limit of \emph{any} generalised non-local FTL particle model of the form~\eqref{eq:sFTL_particle.1}-\eqref{eq:Theta.3} with sufficiently small non-locality is well approximated by the \emph{ARZ-like model}~\eqref{eq:generalised_ARZ} arising for $\lambda\to 0^+$.
\end{quotation}

\section{Numerical tests}
\label{sect:numerics}
In this section, we compare the numerical results produced by the stochastic particle models~\eqref{eq:OV_particle.1}-\eqref{eq:OV_particle.2},~\eqref{eq:FTL_particle.1}-\eqref{eq:FTL_particle.2} and~\eqref{eq:sFTL_particle.1}-\eqref{eq:Theta.3} with those produced by the corresponding hydrodynamic limits~\eqref{eq:1st_order.non-local},~\eqref{eq:ARZ} and~\eqref{eq:generalised_ARZ} in some test cases. The particle models are solved by means of spatially non-local Monte Carlo algorithms, which will be detailed in the next subsections. Conversely, the first order non-local macroscopic model is solved by means of an upwind-type numerical scheme introduced in~\cite{friedrich2018NHM}, see also~\cite{chiarello2021CHAPTER,chiarello2020EJAM,chiarello2019NHM}; while the second order ARZ and ARZ-like models are solved by a splitting of the transport and source contributions with a classical upwind scheme at every time step to discretise the homogeneous system.

In all tests we consider as spatial domain the interval $[-1,\,1]\subset\R$ with periodic boundary conditions and we prescribe as initial conditions the following traffic density:
$$ \rho_0(x)=
	\begin{cases}
		0.8 & \text{if } x<0 \\
		0.2 & \text{if } x>0
	\end{cases} $$
and, whenever needed, the following mean speed:
$$ u_0(x)=
	\begin{cases}
		0.5 & \text{if } x<0 \\
		0.6 & \text{if } x>0.
	\end{cases} $$

\begin{figure}[!t]
\centering
\includegraphics[width=0.6\textwidth]{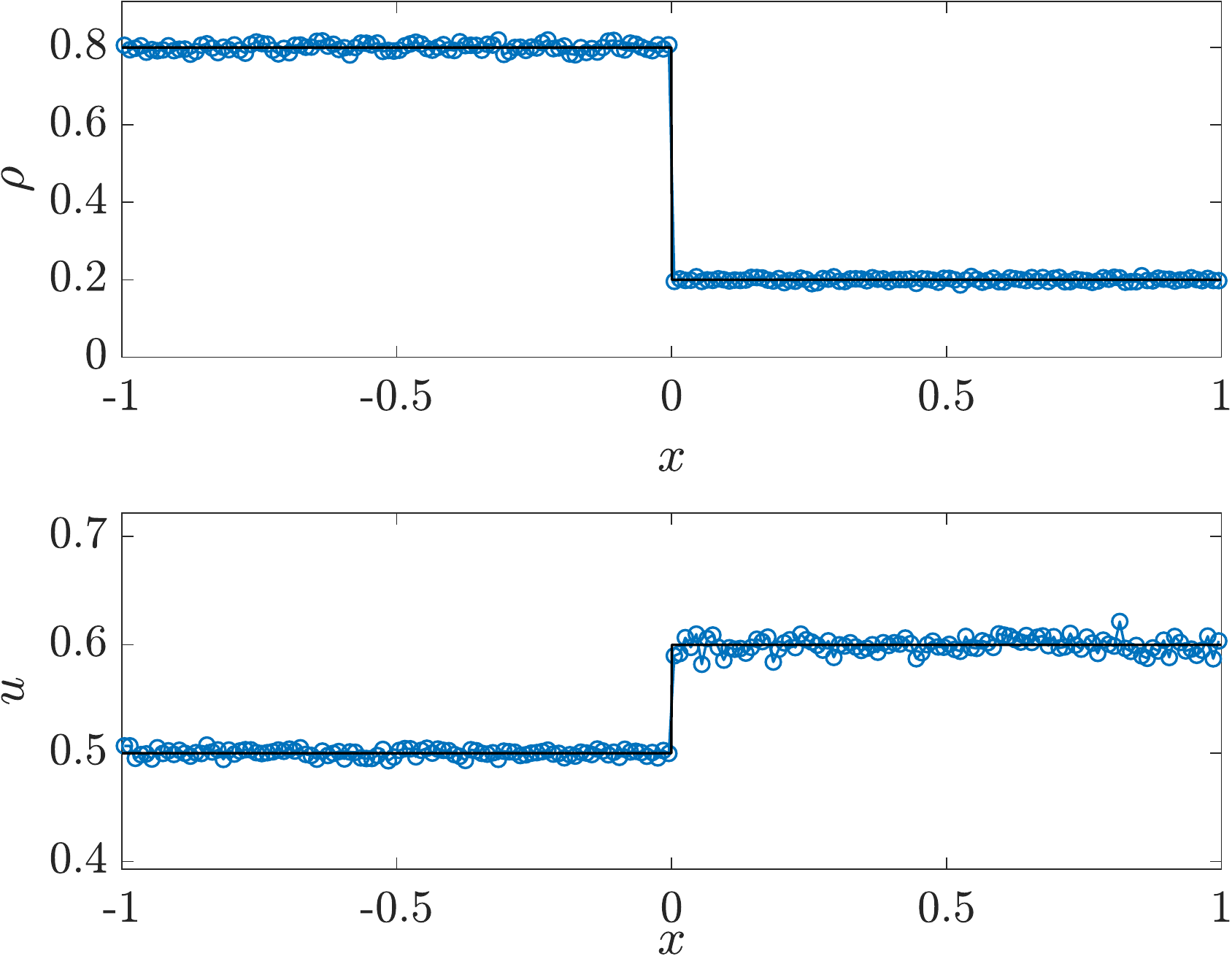}
\caption{Initial condition}
\label{fig:initcond}
\end{figure}

To reproduce the initial traffic density at the particle level, if $N\in\N$ is the total number of particles used in the Monte Carlo algorithm we sample uniformly $0.8N$ particle positions in the interval $[-1,\,0]$ and $0.2N$ particle positions in the interval $[0,\,1]$. Likewise, to reproduce the initial mean speed, for $x<0$ we sample $0.8N$ speed values uniformly distributed in the interval $[0,\,1]$, so that their (theoretical) mean is $0.5$; while for $x>0$ we sample $0.2N$ speed values uniformly distributed in the interval $[0.2,\,1]$, so that their (theoretical) mean is $0.6$. We refer the reader to~\cite{chiarello2021IJNM} for more details about the procedure to link particles and hydrodynamic variables in the implementation of numerical algorithms. Figure~\ref{fig:initcond} depicts the initial condition just discussed.

Moreover, we fix the following interaction kernel:
$$ B(y)=\left(1-\frac{y}{\eta}\right)\chi(0\leq y\leq\eta), $$
which complies with Assumption~\ref{ass:B} and decreases linearly from $1$ to $0$ in the interval $[0,\,\eta]$ (hence $\max_{y\in [0,\,\eta]}B(y)=1$). Such a kernel models a stronger influence of closer vehicles and a weaker influence of farther vehicles and is such that
$$ \cB_0=\frac{\eta}{2}, \qquad \cB_1=\frac{\eta^2}{6}. $$

\subsection{Optimal speed dynamics}
\begin{algorithm}[!t]
	\caption{Non-local Monte Carlo algorithm for the particle model~\eqref{eq:OV_particle.1}-\eqref{eq:OV_particle.2}}
	\label{alg:MC}
	Fix $\Delta{t}=\frac{\eps}{\max_{y\in [0,\,\eta]}B(y)}$\;
	Fix $\Delta{x}\leq\eta$ and consider a space mesh $\{E_h\}_{h\in\Z}$ of pairwise disjoint cells with $\abs{E_h}=\Delta{x}$ and $\cup_{h\in\Z}E_h=\R$\;
	\For{$n=0,\,1,\,2,\,\dots$}{
		\For{$h\in\Z$}{
			Find the particles belonging to the cell $E_h$. Let $\mathcal{E}^n_h:=\{i\in\{1,\dots,N\}:x^n_i\in E_h\}$\;
			\For{$i\in\mathcal{E}^n_h$}{
				Sample randomly $k\in\Z$ such that $h\leq k\leq h+\lfloor{\frac{\eta}{\Delta{x}}}\rfloor$\;
				Sample randomly $j\in\mathcal{E}^n_k$\;
				Let $\rho^n_k$ be an approximation of the traffic density $\rho$ in $E_k$ at time $t_n:=n\Delta{t}$\;
				Sample $\Theta\sim\operatorname{Bernoulli}(B(x^n_j-x^n_i)\frac{\Delta{t}}{\eps})$, $\Theta=\theta\in\{0,\,1\}$\;
				Update $v^n_i$ to $v^{n+1}_i$ according to~\eqref{eq:OV_particle.1}: $v^{n+1}_i=v^n_i+\theta a(\cV(\rho^n_k)-v^n_i)$\;
				}
		}
	Update the particle positions according to~\eqref{eq:OV_particle.1}: $x^{n+1}_i=x^n_i+v^{n+1}_i\Delta{t}$, $i\in\{1,\dots,N\}$\;
}
\end{algorithm}

We begin by considering the stochastic particle model~\eqref{eq:OV_particle.1}-\eqref{eq:OV_particle.2}, to which there corresponds the first order non-local macroscopic model~\eqref{eq:1st_order.non-local}. In particular, we choose the following optimal speed function:
\begin{equation}
	\cV(\rho)=\frac{1}{\tanh{(1)}}\tanh{\left(\frac{1}{1+\rho}\right)},
	\label{eq:V1}
\end{equation}
as a prototype inspired by~\cite{bando1995PRE} which complies with Assumption~\ref{ass:V}. In Algorithm~\ref{alg:MC} we detail the implementation of the Monte Carlo scheme that we use to approach numerically the particle model.

\begin{figure}[!t]
\centering
\includegraphics[width=0.325\textwidth]{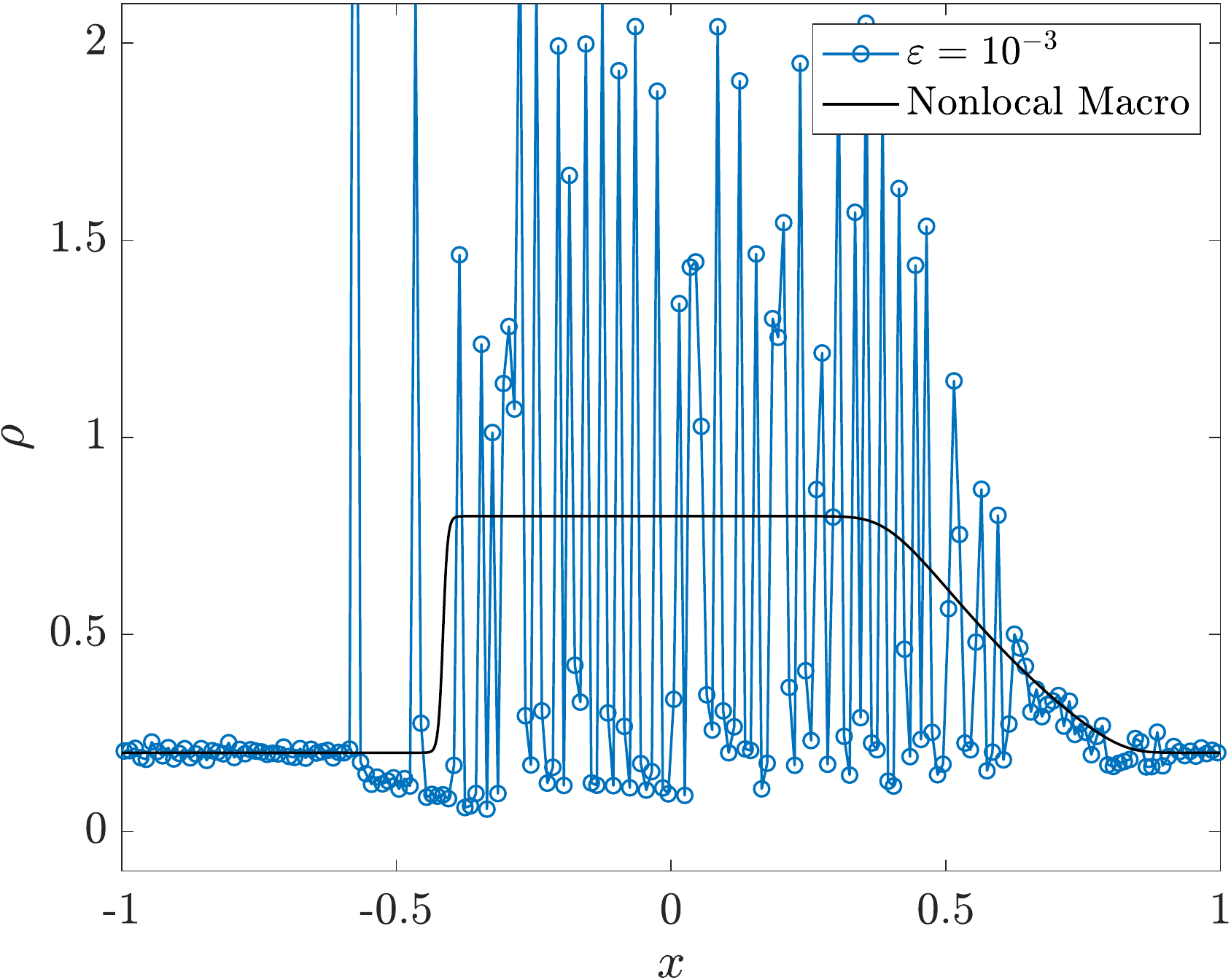}
\includegraphics[width=0.325\textwidth]{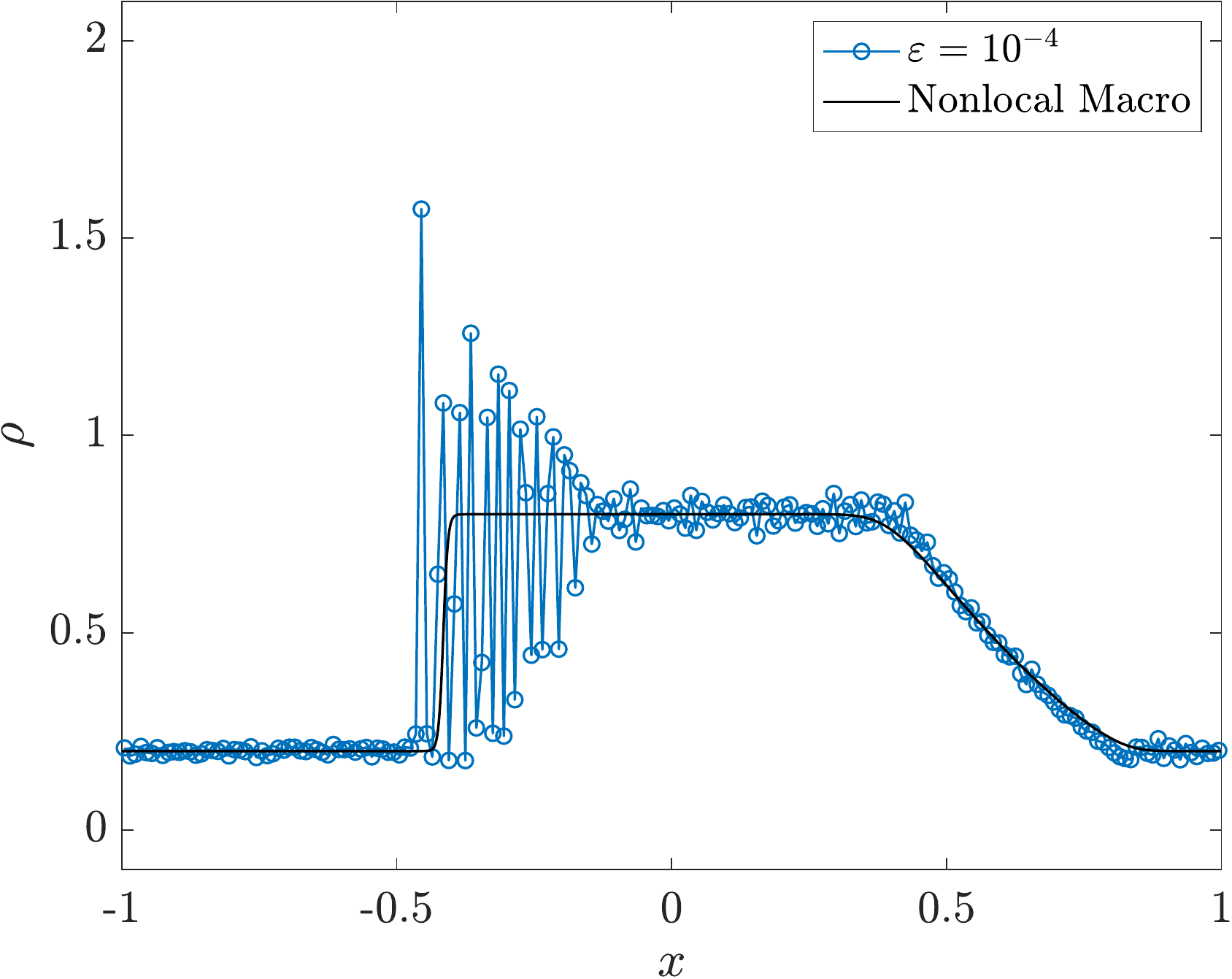}
\includegraphics[width=0.325\textwidth]{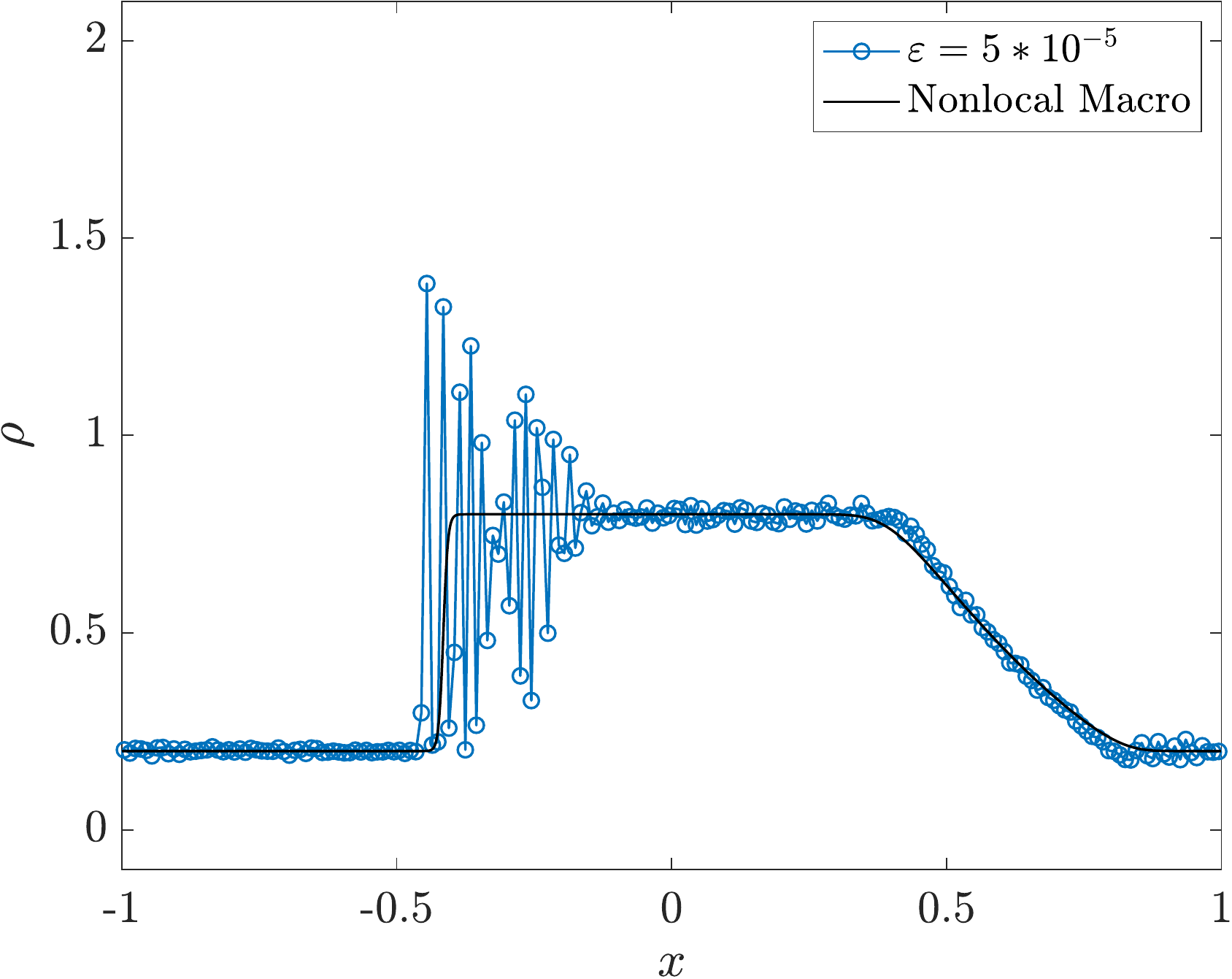}
\caption{Solution of the particle model~\eqref{eq:OV_particle.1}-\eqref{eq:OV_particle.2} (markers) obtained with $N=10^6$ particles and $\Delta{x}=10^{-2},$ and of the corresponding first order non-local hydrodynamic model~\eqref{eq:1st_order.non-local} (solid line) at the computational time $t=1$ for fixed $\eta=10^{-2}$, $a=0.5,\,\Delta{x}=10^{-3}$ and scaling parameter $\eps$ decreasing from $10^{-3}$ to $5\cdot 10^{-5}$ (left to right)}
\label{fig:1st_order.decr_eps}
\end{figure}

Figure~\ref{fig:1st_order.decr_eps} shows that the density profiles of the solutions produced by the particle model and by the non-local hydrodynamic model tend to coincide substantially as the scaling parameter $\eps$ decreases from $10^{-3}$ to $5\cdot 10^{-5}$. This correctly reproduces the hydrodynamic limit anticipated by the theory. We notice that for $\eps=5\cdot 10^{-5}$ some oscillations still appear in the particle solution in correspondence of the rear shock wave exhibited by the hydrodynamic solution. These oscillations, which are the microscopic counterpart of the sharp localised variation in the mass of vehicles, may be smeared out by further reducing the scaling parameter $\eps$, however at the price of an increased computational cost of the Monte Carlo algorithm, where $\Delta{t}=O(\eps)$ (cf. Algorithm~\ref{alg:MC}).

\begin{figure}[!t]
\centering
\includegraphics[width=0.4\textwidth]{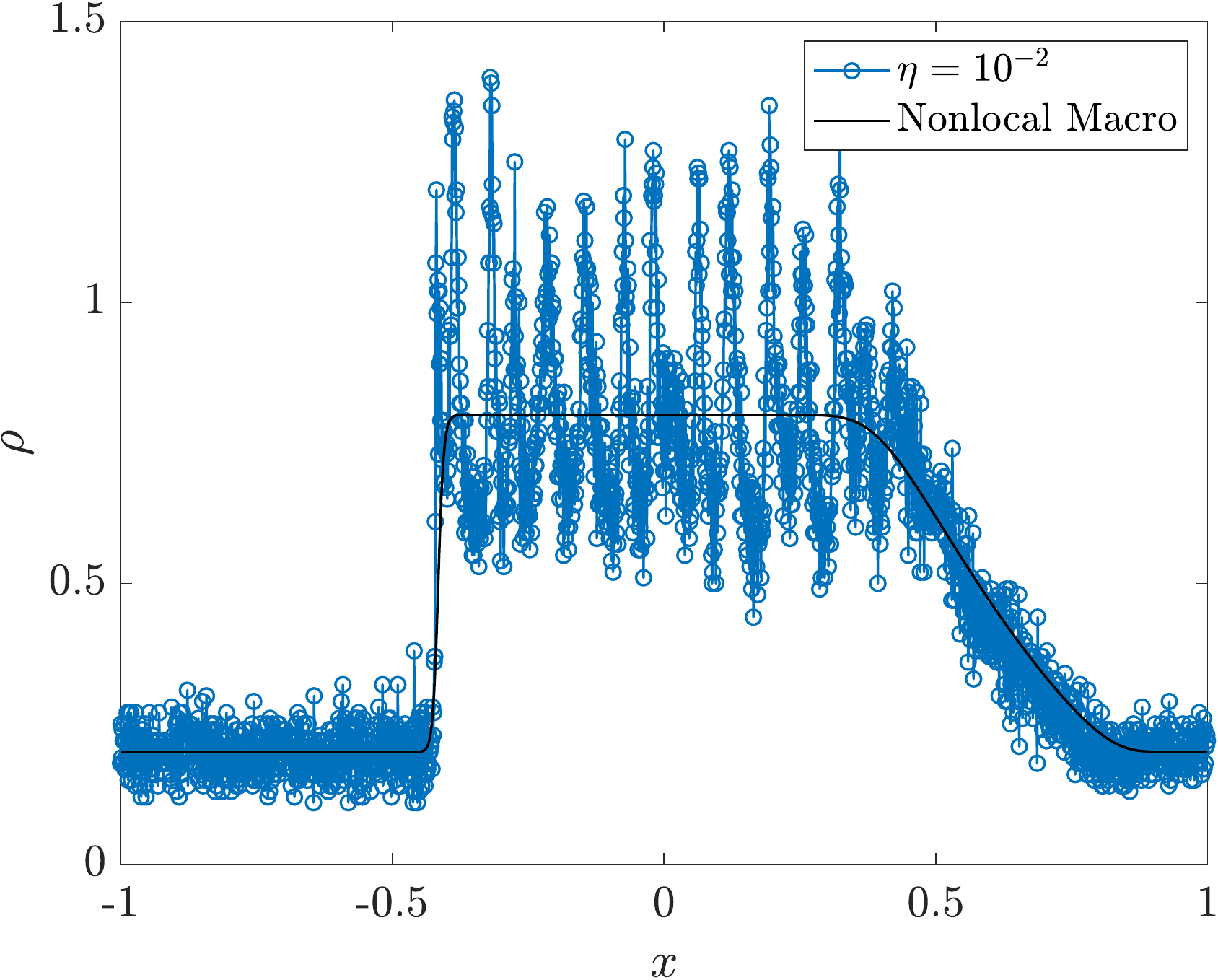}
\includegraphics[width=0.4\textwidth]{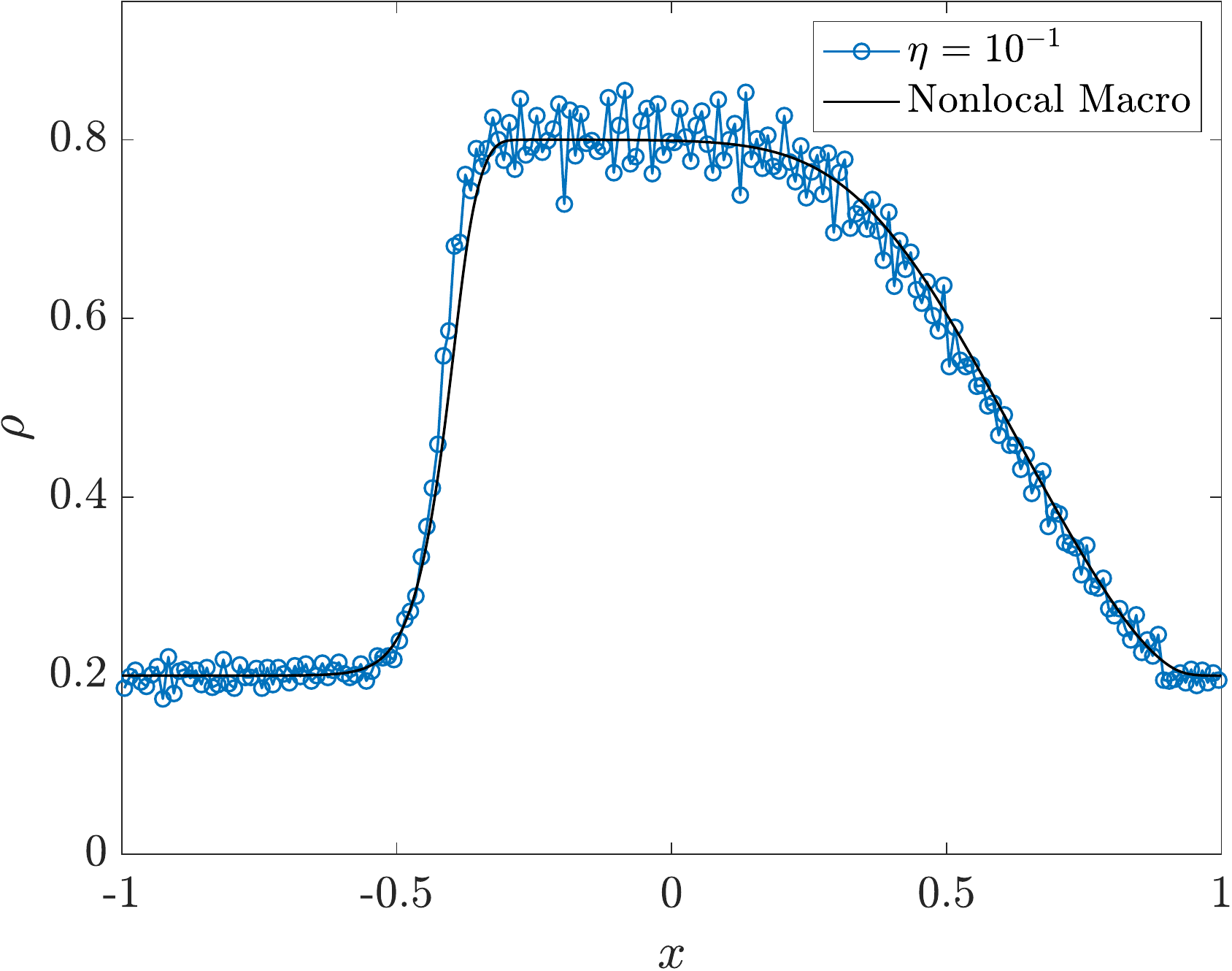} 
\caption{Solution of the particle model~\eqref{eq:OV_particle.1}-\eqref{eq:OV_particle.2} (markers) obtained with $N=10^5$ particles, $\Delta{x}=10^{-3}$ at left and $\Delta{x}=10^{-2}$ at right, and of the corresponding first order non-local hydrodynamic model~\eqref{eq:1st_order.non-local} (solid line) at the computational time $t=1$ for fixed $\eps=10^{-2}$, $a=0.5,\,\Delta{x}=10^{-3}$ and size $\eta$ of the support of $B$ increasing from $10^{-2}$ to $10^{-1}$ (left to right)}
\label{fig:1st_order.decr_eta}
\end{figure}

Figure~\ref{fig:1st_order.decr_eta} shows instead the effect of the size $\eta$ of the support of the interaction kernel $B$ on the agreement between the particle and hydrodynamic density profiles for fixed scaling parameter $\eps$. We clearly notice that the larger the support the more damped the oscillations of the particle model about the average macroscopic solution even for a scaling parameter as moderately small as $\eps=10^{-2}$.

\begin{figure}[!t]
\centering
\includegraphics[scale=0.35]{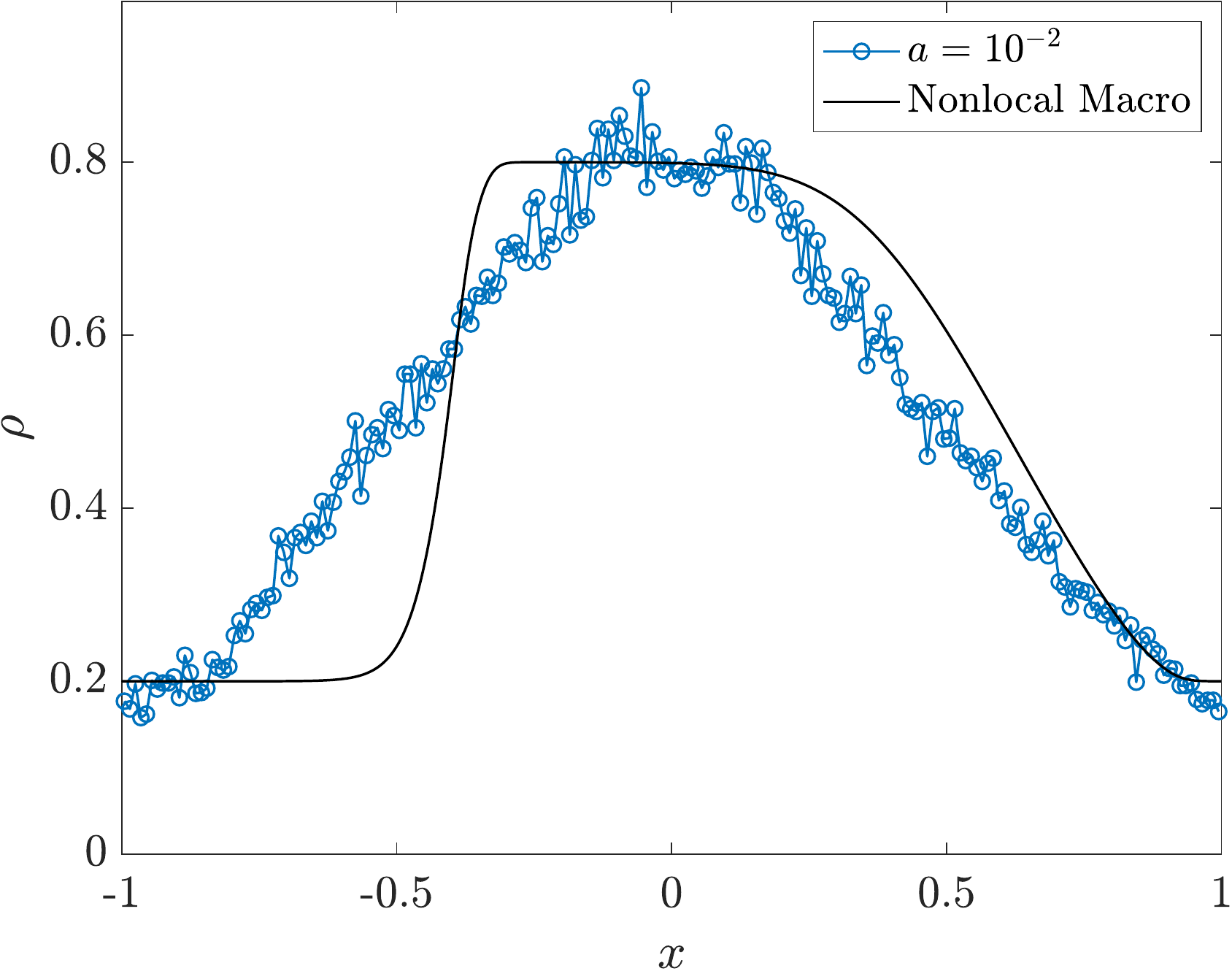}
\includegraphics[scale=0.35]{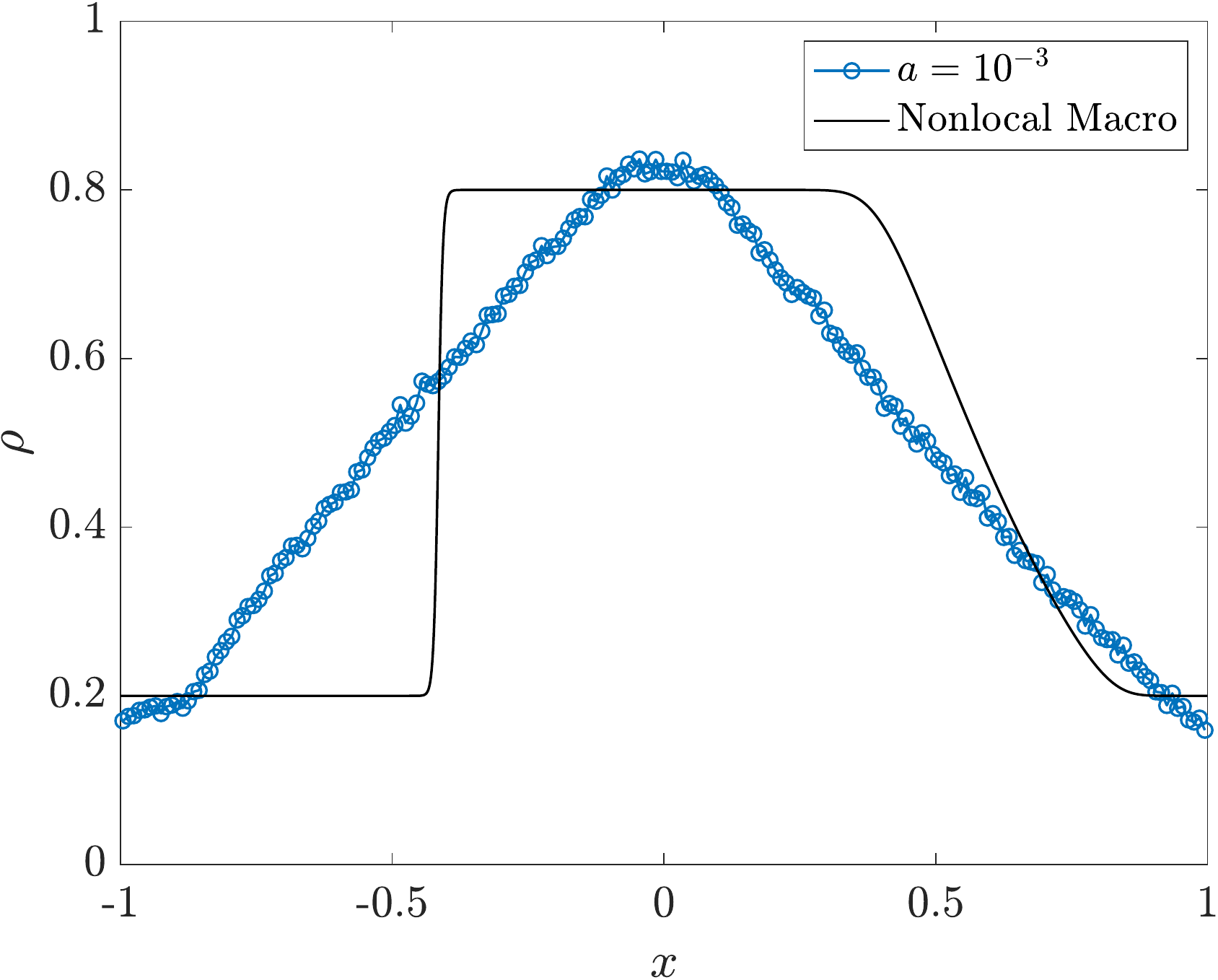}
\caption{Solution of the particle model~\eqref{eq:OV_particle.1}-\eqref{eq:OV_particle.2} (markers) with $\Delta{x}=10^{-2}$ and of the first order non-local hydrodynamic model~\eqref{eq:1st_order.non-local} (solid line) at the computational time $t=1$ with $\Delta{x}=10^{-3}$ for the following sets of parameters: (left) $a=\eps=10^{-2}$, $\eta=10^{-1}$, $N=10^5$; (right) $a=\eps=10^{-3}$, $\eta=10^{-2}$, $N=10^6$}
\label{fig:1st_order.small_a}
\end{figure}

Now we investigate how the relaxation parameter $a$ affects the reliability of the aggregate description provided by the non-local hydrodynamic model~\eqref{eq:1st_order.non-local} with respect to the original particle model~\eqref{eq:OV_particle.1}-\eqref{eq:OV_particle.2}. This is especially important in practical situations where the scaling parameter $\eps$ might not be taken arbitrarily small, such as e.g., in numerical simulations. Although the limit model~\eqref{eq:1st_order.non-local} is unaffected by the value of $a$, for finite $\eps>0$ one cannot rely on the exact limit condition~\eqref{eq:local.Maxwellian} and has instead to settle for an approximate condition of the form $Q(f^\eps,f^\eps)\approx 0$, which implies (cf.~\eqref{eq:int.vQ})
$$ u^\eps\approx\frac{\tilde{B}\ast(\rho^\eps\cV(\rho^\eps))}{\tilde{B}\ast\rho^\eps}. $$
One may expect such an approximation to hold when $\eps$ is sufficiently small. In fact, the validity of the approximation does not depend on $\eps$ alone but on ratio $a/\eps$. This becomes apparent multiplying~\eqref{eq:kinetic.strong.eps-1st.ord} by $v$ and integrating with respect to $v$, which, taking~\eqref{eq:int.vQ} into account, yields
$$ \partial_t\int_0^1vf^\eps\,dv+\partial_x\int_0^1v^2f^\eps\,dv=\frac{a}{2\eps}\rho^\eps\left(\tilde{B}\ast(\rho^\eps\cV(\rho^\eps))-(\tilde{B}\ast\rho^\eps)u^\eps\right). $$
If $a/\eps$ is sufficiently large, i.e. $a\gg\eps$, then the left-hand side of this equation is formally negligible with respect to the right-hand side, which makes the approximation above for $u^\eps$ numerically acceptable. Conversely, if $a/\eps$ is not large enough, because either $a\sim\eps$ or $a\ll\eps$, then the left-hand side of the previous equation is not actually negligible with respect to the right-hand side. In this case, the approximation for $u^\eps$ may not be fully justified and the hydrodynamic description~\eqref{eq:1st_order.non-local} may not provide a reliable description of the particle dynamics~\eqref{eq:OV_particle.1}-\eqref{eq:OV_particle.2} for the given value of $\eps$. Figure~\ref{fig:1st_order.small_a} illustrates two examples of disagreement between the particle and the macroscopic solutions due to $a=\eps$ for different sets of the other parameters.

\subsection{Follow-The-Leader dynamics}
\begin{algorithm}[!t]
	\caption{Non-local Monte Carlo algorithm for the particle model~\eqref{eq:FTL_particle.1}-\eqref{eq:FTL_particle.2}}
	\label{alg:MC2}
	Fix	$\Delta{t}=\frac{\eps}{\max_{y\in [0,\,\eta]}B(y)}$\;
	Fix\footnotemark $\Delta{x}=\eta$ and consider a space mesh $\{E_h\}_{h\in\Z}$ of pairwise disjoint cells with $\abs{E_h}=\Delta{x}$ and $\cup_{h\in\Z}E_h=\R$\;
	\For{$n=0,\,1,\,2,\,\dots$}{
		\For{$h\in\Z$}{
			Find the particles belonging to the cell $E_h$. Let $\mathcal{E}^n_h:=\{i\in\{1,\dots,N\}:x^n_i\in E_h\}$\;
			\For{$i\in\mathcal{E}^n_h$}{
				Sample randomly $j\in\mathcal{E}^n_h$\;
				Sample $\Theta\sim\operatorname{Bernoulli}(B(x^n_j-x^n_i)\frac{\Delta{t}}{\eps})$, $\Theta=\theta\in\{0,\,1\}$\;
				Update $v^n_i$ to $v^{n+1}_i$ according to~\eqref{eq:FTL_particle.1}: $v^{n+1}_i=v^n_i+\theta\lambda(v^n_j-v^n_i)$\;
			}
		}
		Update particle positions according to~\eqref{eq:FTL_particle.1}: $x^{n+1}_i=x^n_i+v^{n+1}_i\Delta{t}$, $i\in\{1,\dots,N\}$\;
	}
\end{algorithm}
\footnotetext{In second order models, $\eta$ has to be small for the theory developed in the previous sections to hold. For this reason, unlike Algorithm~\ref{alg:MC}, in Algorithm~\ref{alg:MC2} it is fair to take the mesh size $\Delta{x}$ directly equal to $\eta$.}

We consider now the stochastic FTL particle model~\eqref{eq:FTL_particle.1}-\eqref{eq:FTL_particle.2}, to which there corresponds the second order macroscopic ARZ model~\eqref{eq:ARZ} at least in the regime of small non-locality, cf.~\eqref{eq:eta.small}. Algorithm~\ref{alg:MC2} reports the implementation of the Monte Carlo scheme that we use in this case to solve the particle model numerically.

\begin{figure}[!t]
\centering
\includegraphics[width=0.460\textwidth]{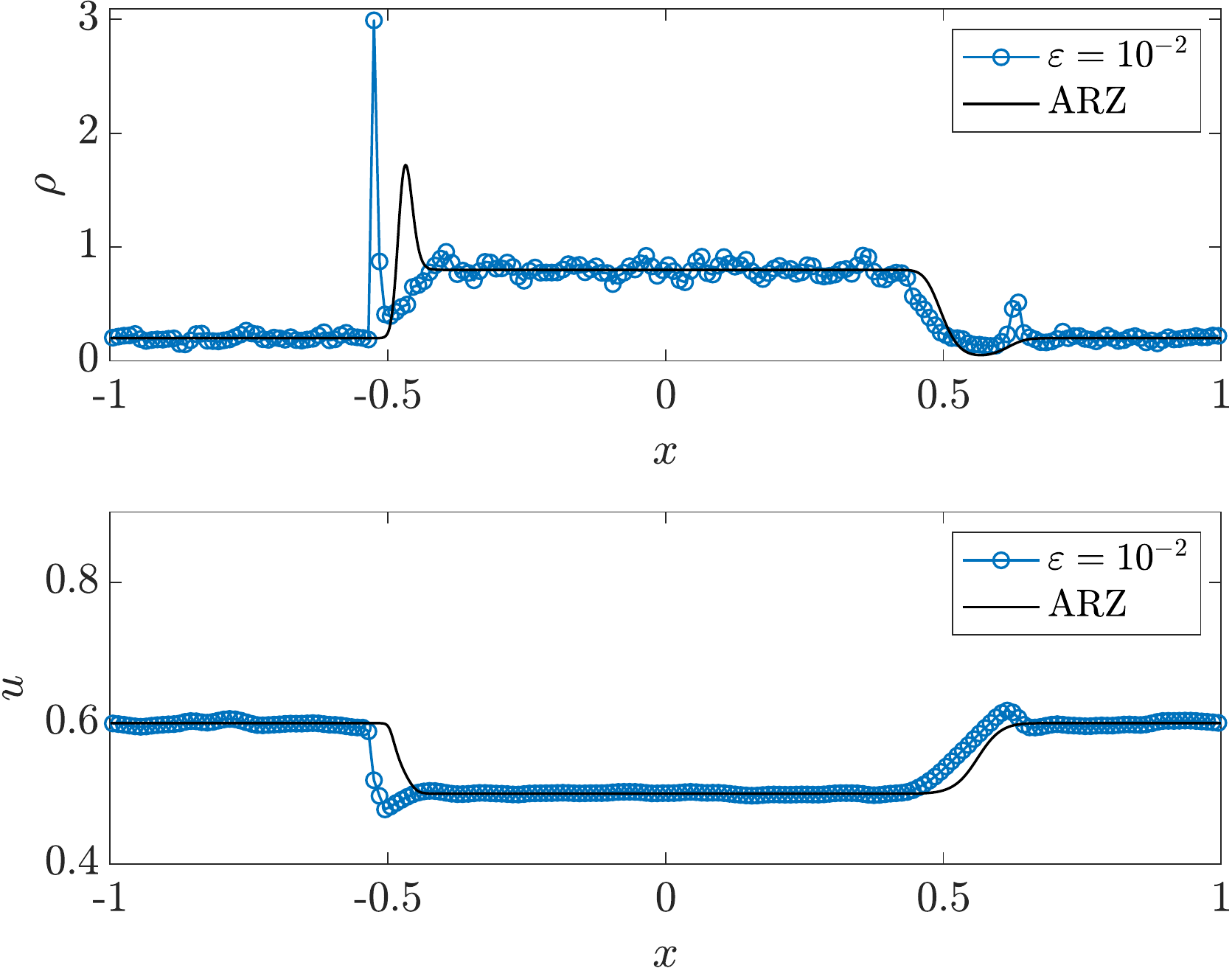}
\includegraphics[width=0.450\textwidth]{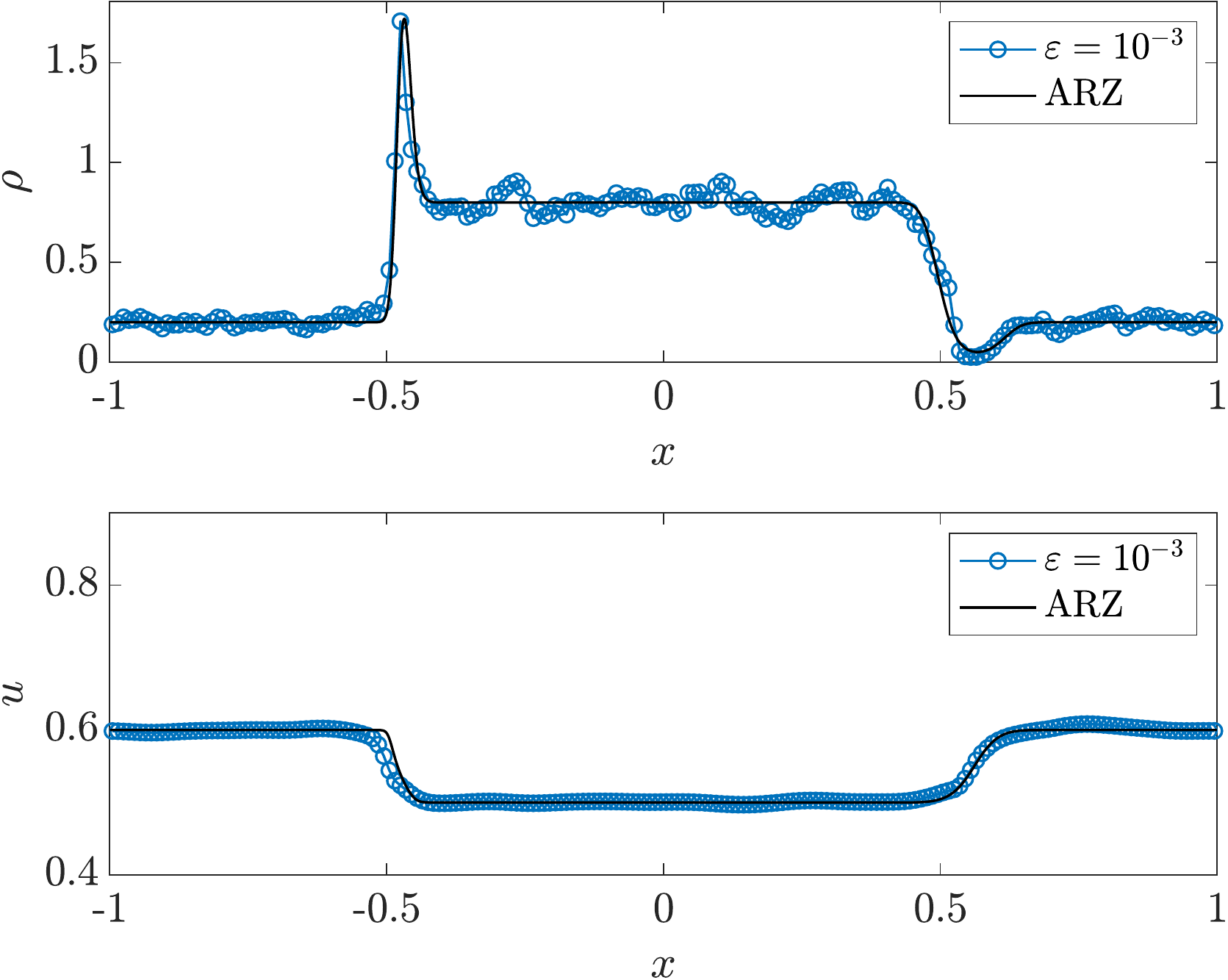}
\caption{Solution of the particle model~\eqref{eq:FTL_particle.1}-\eqref{eq:FTL_particle.2} (markers) with $N=10^6$ particles and $\Delta{x}=10^{-2},$ and of the ARZ model~\eqref{eq:ARZ} (solid line) at the computational time $t=1$, $\Delta{x}=10^{-3}$ for $\eta=10^{-2}$ (\textit{small} support of the interaction kernel) and $\lambda=0.5$. Left column: $\eps=10^{-2}$; right column: $\eps=10^{-3}$}
\label{fig:2nd_order.small_eta}
\end{figure}

Figure~\ref{fig:2nd_order.small_eta} shows that for a sufficiently small support of the interaction kernel, in particular $\eta=10^{-2}$, both the traffic density $\rho$ and the mean speed $u$ computed out of the particle model tend to be well reproduced by the ARZ model when the scaling parameter $\eps$ decreases from $10^{-2}$ to $10^{-3}$. This is in agreement with the hydrodynamic limit predicted by the theory.

\begin{figure}[!t]
\centering
\includegraphics[width=0.46\textwidth]{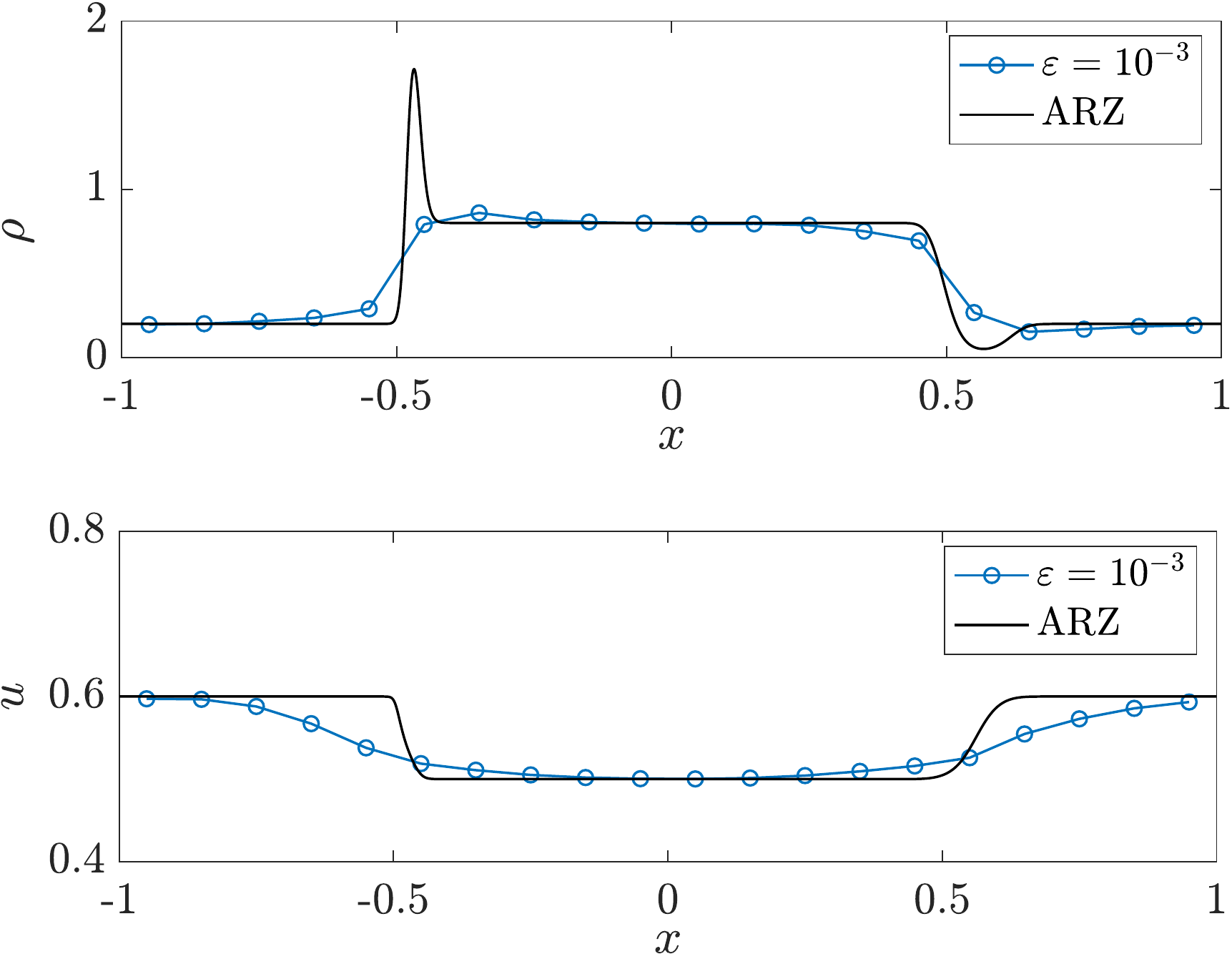}
\includegraphics[width=0.45\textwidth]{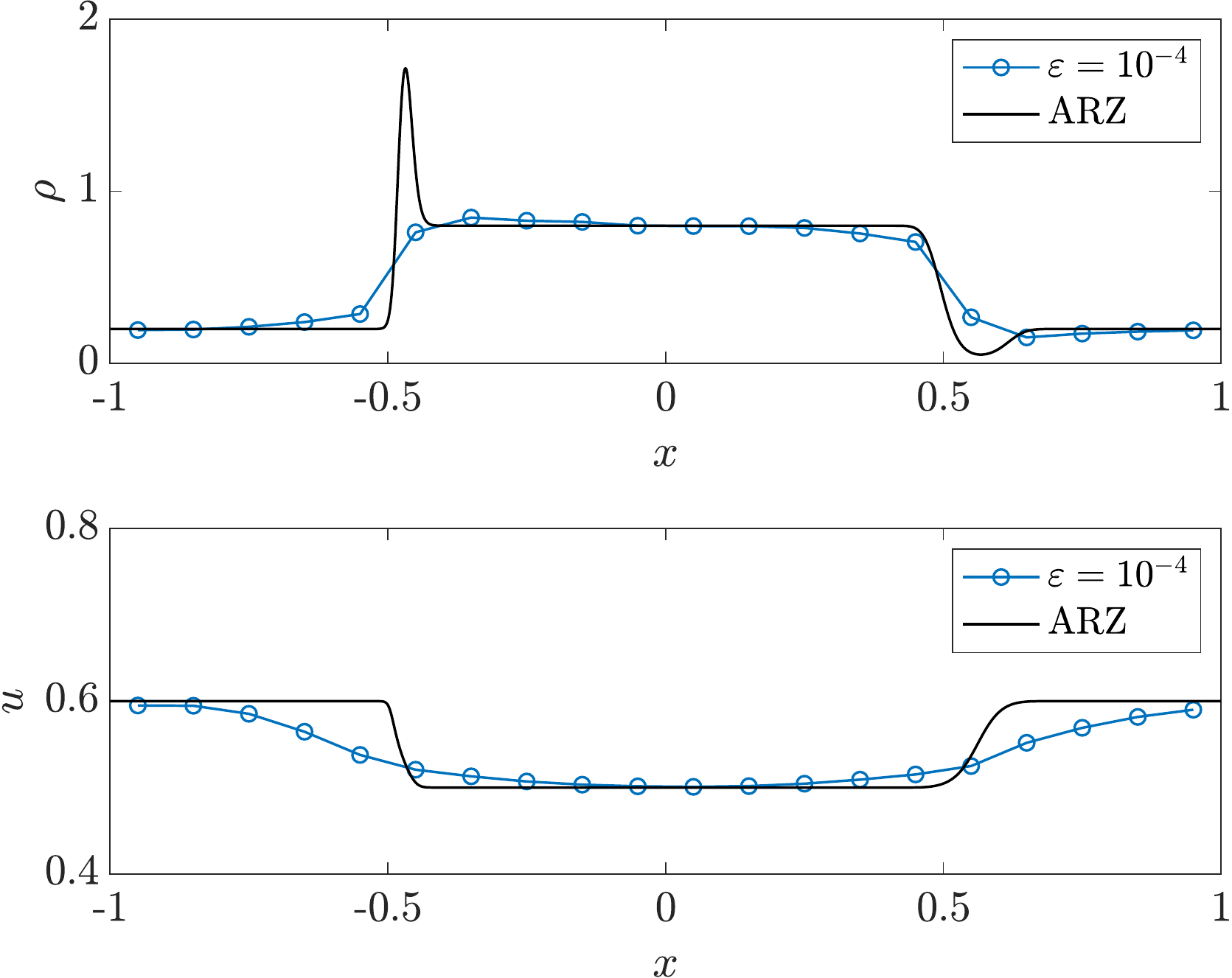}
\caption{Solution of the particle model~\eqref{eq:FTL_particle.1}-\eqref{eq:FTL_particle.2} (markers) with $N=10^6$ particles and $\Delta{x}=10^{-2},$ and of the ARZ model~\eqref{eq:ARZ} (solid line) at the computational time $t=1$, $\Delta{x}=10^{-3}$ for $\eta=10^{-1}$ (\textit{large} support of the interaction kernel) and $\lambda=0.5$. Left column: $\eps=10^{-3}$; right column: $\eps=10^{-4}$}
\label{fig:2nd_order.large_eta}
\end{figure}

Figure~\ref{fig:2nd_order.large_eta} shows instead that if the support of the interaction kernel is not small, in this case $\eta=10^{-1}$, then there is no agreement between the solutions of the particle and the ARZ models even for values of the scaling parameter much smaller than before, in this case $\eps=10^{-3},\,10^{-4}$. In particular, we notice that the particle solutions obtained with the two tested values of $\eps$ are virtually the same, thereby suggesting that the hydrodynamic limit has been substantially reached numerically. The fact that such solutions do not coincide with the one produced by the ARZ model proves that, in the regime of large $\eta$, the ARZ model is not the macroscopic counterpart of the particle system. Indeed, that the approximation~\eqref{eq:f.approx}, which the hydrodynamic limit leading to the ARZ model is based on, is not valid in the considered regime.

\begin{figure}[!t]
\centering
\includegraphics[width=0.325\textwidth]{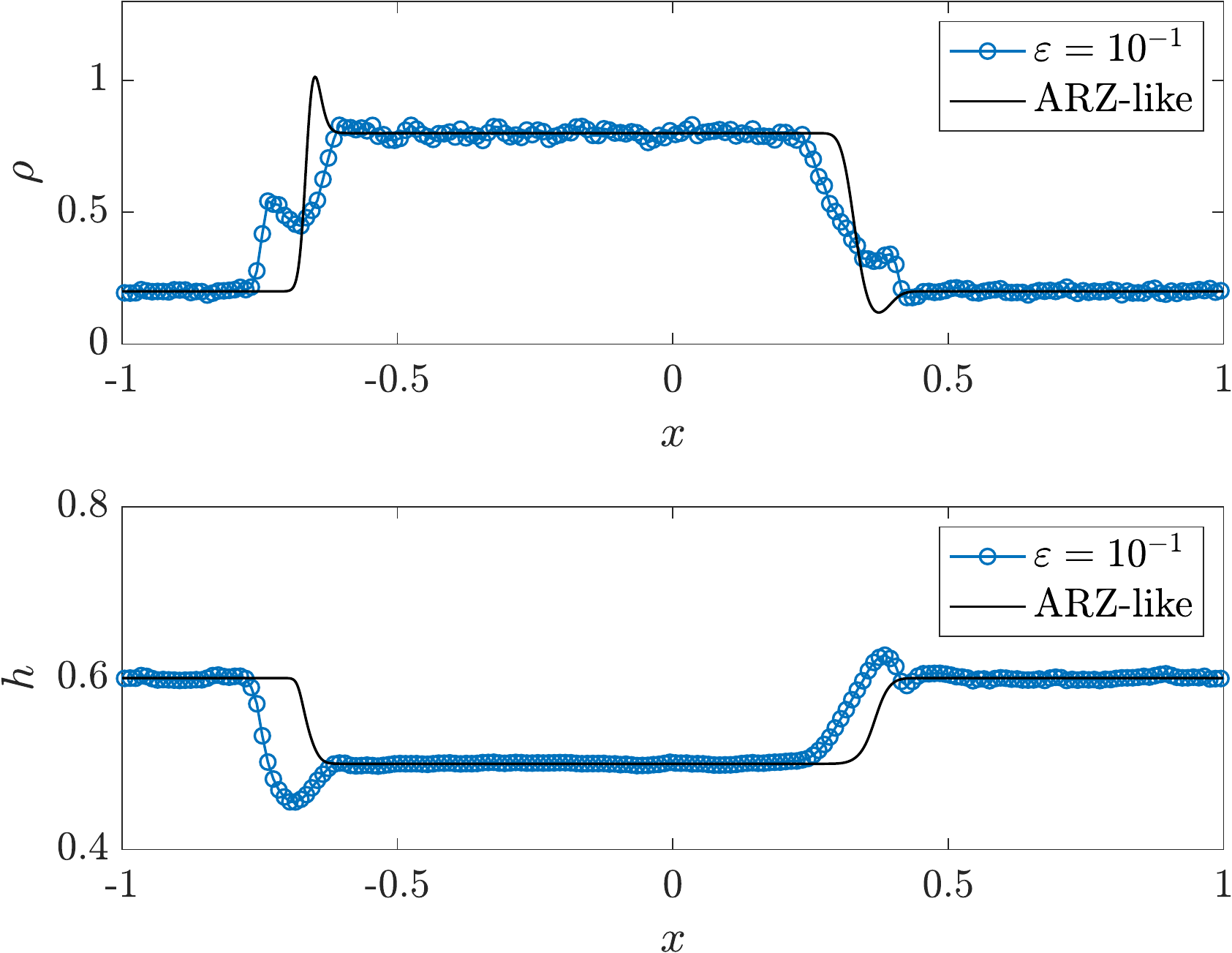} 
\includegraphics[width=0.325\textwidth]{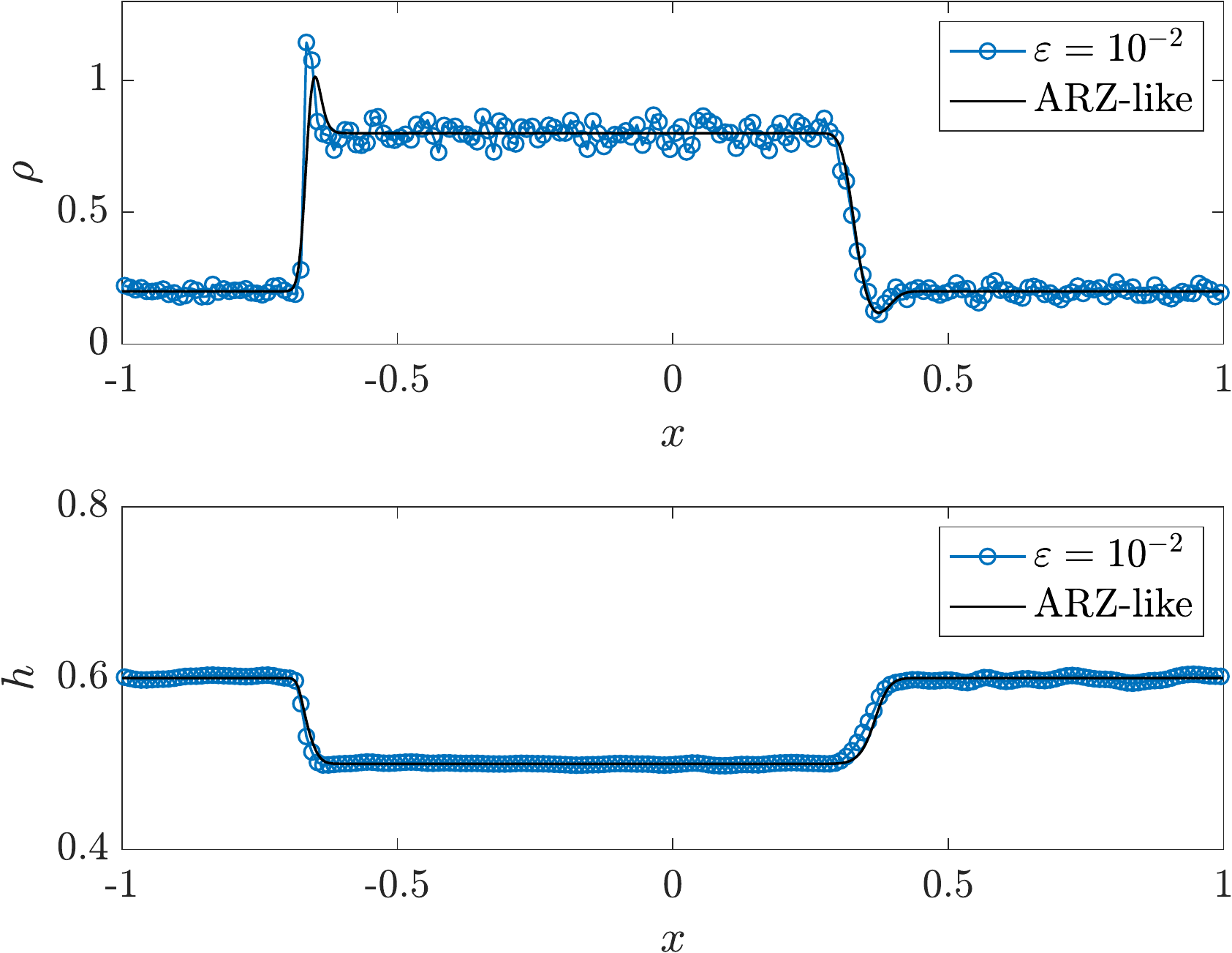} 
\includegraphics[width=0.325\textwidth]{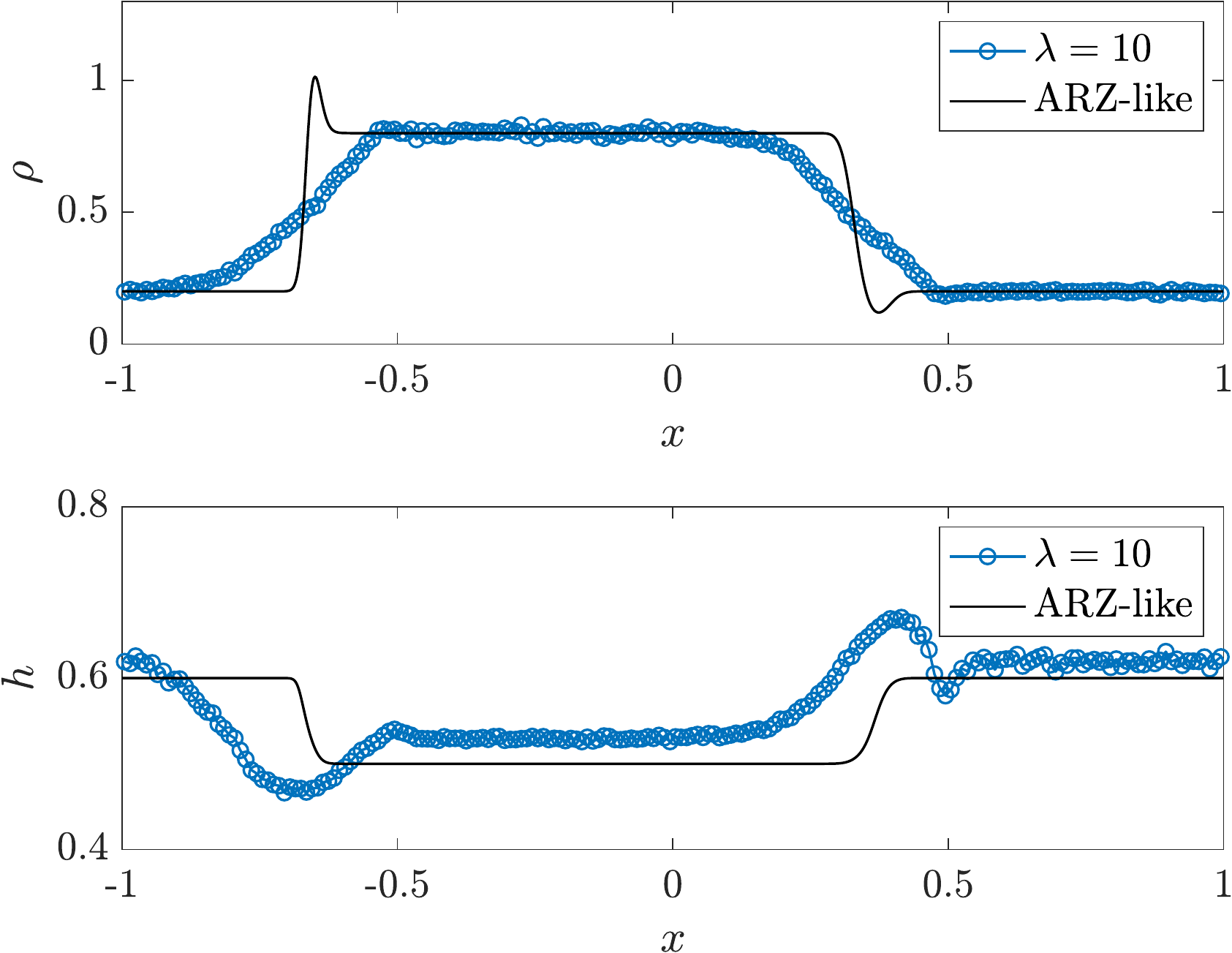} 
\caption{Solution of the particle model~\eqref{eq:sFTL_particle.1}-\eqref{eq:Theta.3} (markers) with $N=10^6$ particles and $\Delta{x}=10^{-2},$ and of the ARZ-like model~\eqref{eq:generalised_ARZ} (solid line) at the computational time $t=1$, $\Delta{x}=10^{-3}$ for $\eta=10^{-2}.$  Left column: $\eps=10^{-1},\, \lambda=10^{-1}$; middle column: $\eps=10^{-2},\, \lambda=10^{-1};$ right column: $\eps=10^{-2},\, \lambda=10$}
\label{fig:2nd_order.h.small_eta}
\end{figure}

Finally, Figure~\ref{fig:2nd_order.h.small_eta} shows a comparison between the solutions produced by the generalised FTL particle model~\eqref{eq:sFTL_particle.1}-\eqref{eq:Theta.3} with
\begin{equation}
	\cV(s)=\frac{s}{1+s} , \qquad \Psi_\lambda(s)=\frac{\lambda s}{1+\lambda s}
	\label{eq:V2}
\end{equation}
and the generalised ARZ model~\eqref{eq:generalised_ARZ} in the regime $\eta\ll 1$. 
We solve the particle model by an algorithm analogous to Algorithm~\ref{alg:MC2} with due modifications in the interaction rules, cf.~\eqref{eq:sFTL_particle.1}-\eqref{eq:sFTL_particle.2}, and the interaction kernel, cf.~\eqref{eq:Theta.3}. Furthermore, as initial condition for the headway $h$ we prescribe the function
$$ h_0(x)=
	\begin{cases}
		0.5 \quad x<0 \\
		0.6 \quad x>0.
	\end{cases} $$
Also in this case we observe (Figure~\ref{fig:2nd_order.h.small_eta} -- left and middle columns) that, consistently with the theory developed in Section~\ref{sect:gen_FTL}, the particle solution approaches the macroscopic solution as $\eps$ decreases from $10^{-1}$ to $10^{-2}$, provided $\lambda$ is sufficiently small ($\lambda=10^{-1}$ in this case). If instead $\lambda$ is not small enough (e.g., $\lambda=10$, Figure~\ref{fig:2nd_order.h.small_eta} -- right column) the particle solution may be affected consistently by both the function $\Psi_\lambda$ used in the interactions~\eqref{eq:sFTL_particle.1} and the cutoff~\eqref{eq:Theta.3}, in such a way that the hydrodynamic limit~\eqref{eq:generalised_ARZ} does not represent accurately the actual macroscopic dynamics.

\begin{figure}[!t]
\centering
\includegraphics[width=0.445\textwidth]{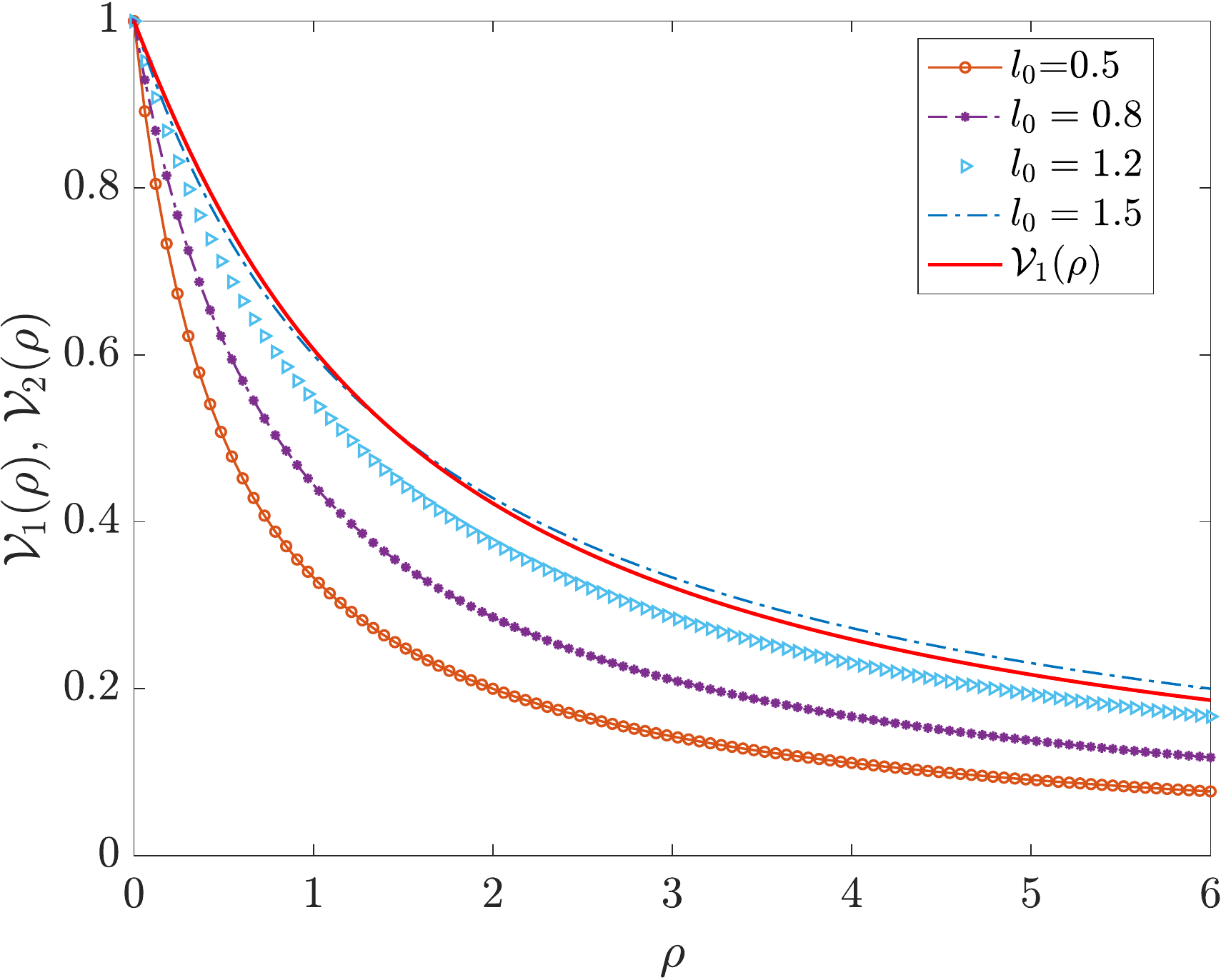}
\caption{Comparison of the speed functions~\eqref{eq:V1},~\eqref{eq:V2} based on the relationship $s=\frac{\ell_0}{\rho}$}
\label{fig:comparison.V}
\end{figure}

\begin{remark}[Speed function comparison]
It may be instructive to compare qualitatively the speed function~\eqref{eq:V2}, say $\cV_2$, used in this numerical test with the one introduced in~\eqref{eq:V1}, say $\cV_1$. We observe that the two functions have a slightly different meaning:~\eqref{eq:V1} is an \textit{optimal speed} towards which vehicle speeds relax depending on the traffic congestion, whereas~\eqref{eq:V2} is the \textit{actual speed} of vehicles expressed in terms of their headway. Nevertheless, invoking the heuristic relationship $s\propto\frac{1}{\rho}$ we may rewrite~\eqref{eq:V2} as
$$ \cV_2(\rho)=\frac{\ell_0}{\rho+\ell_0}, $$
$\ell_0>0$ being the proportionality constant between $s$ and $\frac{1}{\rho}$, which is commonly understood as the characteristic length of a vehicle. Based on this expression of $\cV_2$, Figure~\ref{fig:comparison.V} shows that the trend of the two speed functions is qualitatively the same, hence that they are consistent with one another.
\end{remark}

\subsection{Speeds approaching zero}
\begin{figure}[!t]
\centering
\includegraphics[width=0.45\textwidth]{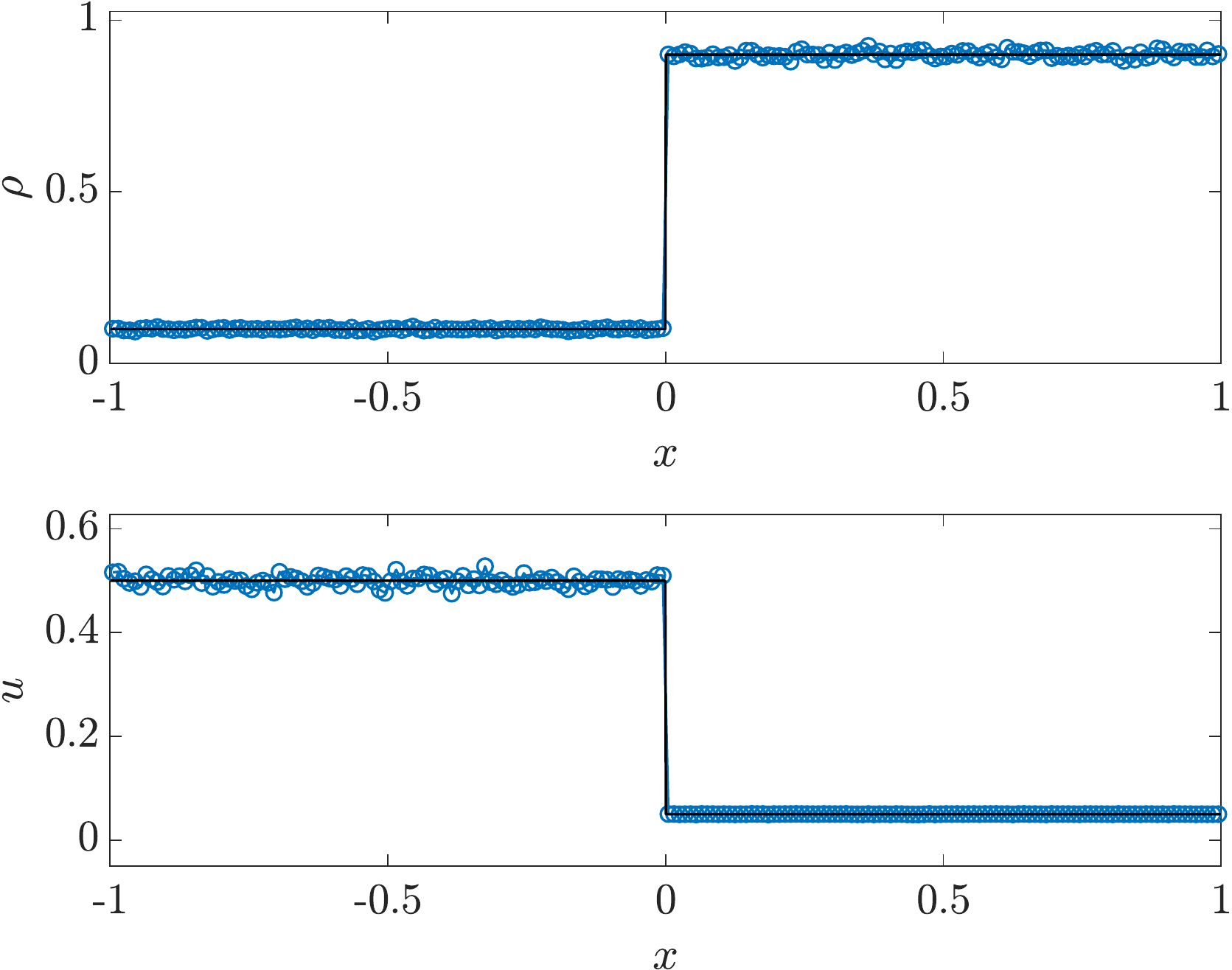}
\includegraphics[width=0.45\textwidth]{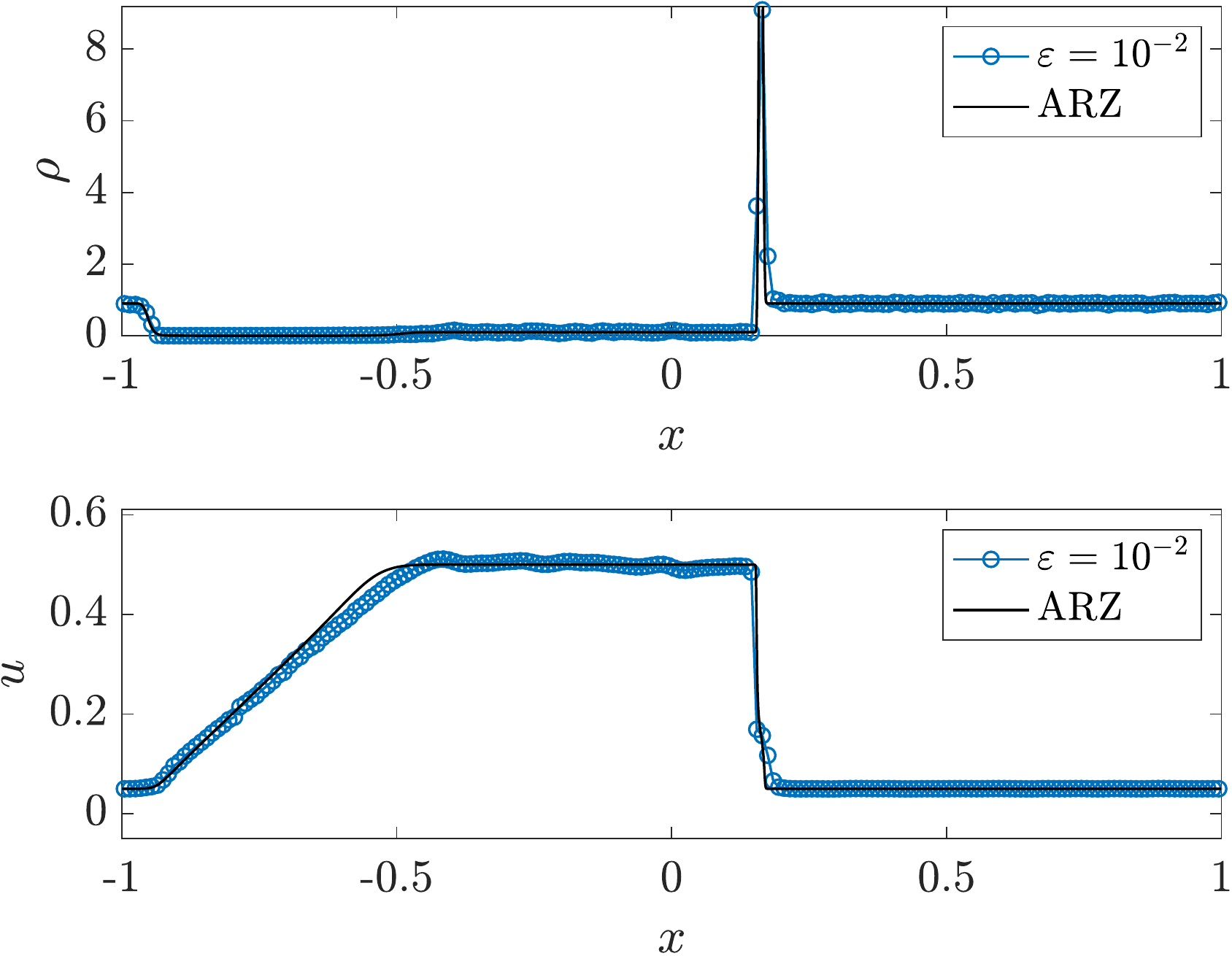} 
\caption{Left: initial condition. Right: solution of the particle model~\eqref{eq:FTL_particle.1}-\eqref{eq:FTL_particle.2} (markers) with $N=10^6$ particles, $\Delta{x}=10^{-2}$ and of the ARZ model~\eqref{eq:ARZ} (solid line) at the computational time $t=1$ with $\Delta{x}=10^{-3}$ for $\eta=10^{-2}$, $\lambda=0.5$ and $\eps=10^{-2}$}
\label{fig:Coda_FTL}
\end{figure}
	
\begin{figure}[!t]
\centering
\includegraphics[width=0.45\textwidth]{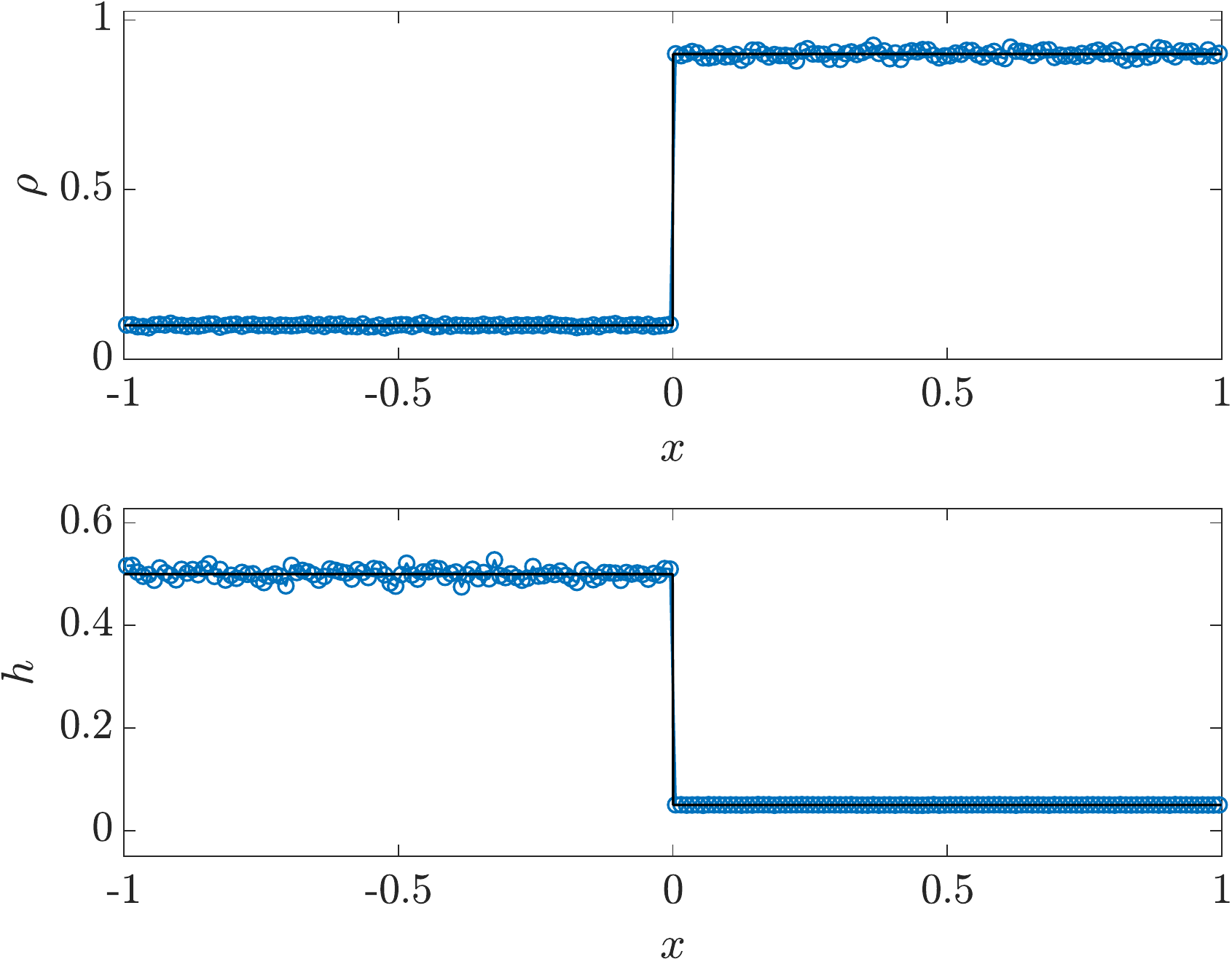}
\includegraphics[width=0.45\textwidth]{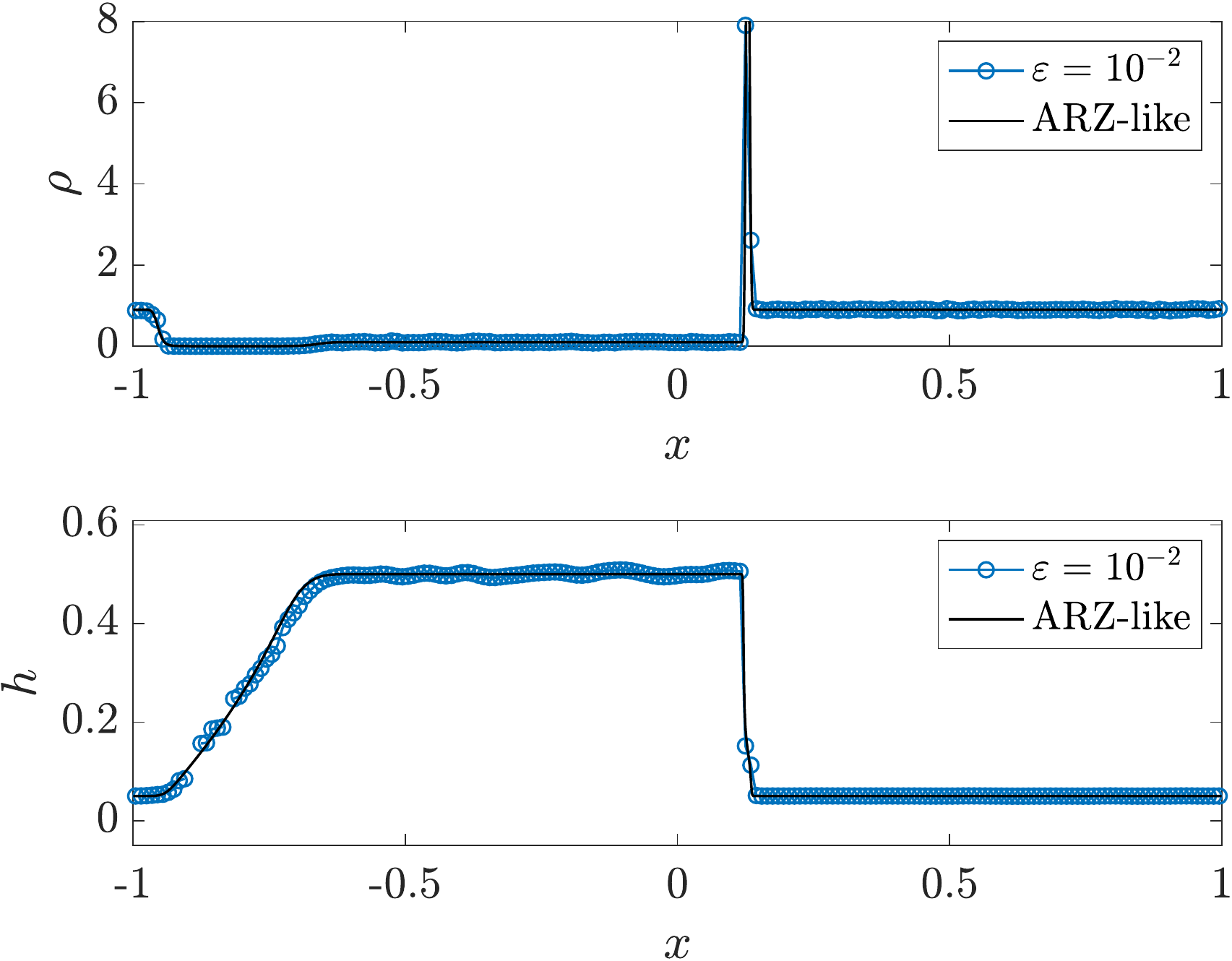} 
\caption{Left: initial condition. Right: solution of the particle model~\eqref{eq:sFTL_particle.1}-\eqref{eq:sFTL_particle.2} (markers) with $N=10^6$ particles, $\Delta{x}=10^{-2}$ and of the ARZ-like model~\eqref{eq:generalised_ARZ} (solid line) at the computational time $t=1$ with $\Delta{x}=10^{-3}$ for $\eta=10^{-2}$, $\lambda=0.5$ and $\eps=10^{-2}$}
\label{fig:Coda_h}
\end{figure}

Finally, we reconsider the two second order models in the case of speeds approaching zero, which for the generalised ARZ-like model means a mean headway approaching zero. The precise initial conditions are plotted in the left panels of Figures~\ref{fig:Coda_FTL},~\ref{fig:Coda_h}. The right panels of the same figures show that, at successive times, the traffic density increases rapidly close to the initial discontinuity, because the speed/headway of vehicles switches suddenly from a high to a low value. We observe that both models do not predict the formation of a queue, i.e. a backward travelling density wave, but rather a pointwise accumulation of the density. At the macroscopic level, this is due to the well-known lack of maximum principle for second order models, which implies no \textit{a priori} upper bound on the vehicle density. However, from the numerical simulations we infer that the macroscopic models are closely reproducing, also in this case, the original particle trends in the proper hydrodynamic regimes. Therefore, we conclude that such a pointwise accumulation of the density is consistent with the (possibly generalised) follow-the-leader dynamics in the regime of small non-locality of the interactions, hence that it is not just an ``analytical drawback'' of the macroscopic description.

\section{Conclusions}
\label{sect:conclusions}
In this paper, we have derived first and second order non-local traffic models as physical limits of fundamental particle dynamics, such as optimal speed and follow-the-leader dynamics. This is, in our view, a first result establishing a structural link between microscopic vehicle dynamics widely recognised as ``first principles'' of traffic and non-local macroscopic descriptions of the flow of vehicles. It is worth stressing that our approach differs from other approaches in the literature, which consider instead the convergence of \textit{ad-hoc} particle discretisations to macroscopic models in the many-particle limit. What these approaches actually prove is, as a matter of fact, that selected particle discretisations may be ``numerically'' consistent with the desired macroscopic models, which remain postulated \textit{a priori}.

The techniques we have used have their roots in the classical collisional kinetic theory revisited in the light of the application to interacting multi-agent systems. More specifically, we have considered hydrodynamic limits of Povzner-type kinetic equations (possibly with cutoff), which are a natural setting to describe non-local vehicle interactions at the mesoscopic scale.

On one hand, we have shown that vehicle interactions based on non-local optimal speed dynamics are well described at the macroscopic scale by a scalar conservation law with non-local flux, whose form however differs slightly from the usual ones assumed in first order models postulated heuristically. On the other hand, we have shown that the macroscopic Aw-Rascle-Zhang traffic model, and possible generalisations of it, emerges as the prototypical hydrodynamic limit of non-local follow-the-leader dynamics with arbitrary (but compactly supported) interaction kernel, as long as the measure of the support of the latter is sufficiently small. These results have also been fully supported by numerical comparisons between the solutions of the particle models and those of  the corresponding hydrodynamic limits.

Further research in this direction may concern the rigorous statement of the formal limits proposed in this paper, as well as the study of the hydrodynamic limit of non-local follow-the-leader vehicle interactions in the case of an interaction kernel with non-small support.

\section*{Acknowledgements}
This work was partially supported  by the Italian Ministry for University and Research (MUR) through the ``Dipartimenti di Eccellenza'' Programme (2018-2022), Department of Mathematical Sciences ``G. L. Lagrange'', Politecnico di Torino (CUP: E11G18000350001) and through the PRIN 2017 project (No. 2017KKJP4X) ``Innovative numerical methods for evolutionary partial differential equations and applications''.

F.A.C. and A.T. are members of GNFM (Gruppo Nazionale per la Fisica Matematica) of INdAM (Istituto Nazionale di Alta Matematica), Italy.
	
\bibliographystyle{plain}
\bibliography{CfTa-kinetic_nonlocal}

\begin{thebibliography}{10}

\bibitem{aw2000SIAP}
A.~Aw and M.~Rascle.
\newblock Resurrection of ``second order'' models of traffic flow.
\newblock {\em SIAM J. Appl. Math.}, 60(3):916--938, 2000.

\bibitem{bando1995PRE}
M.~Bando, K.~Hasebe, A.~Nakayama, A.~Shibata, and Y.~Sugiyama.
\newblock Dynamical model of traffic congestion and numerical simulation.
\newblock {\em Phys. Rev. E}, 51(2):1035--1042, 1995.

\bibitem{blandin2016NM}
S.~Blandin and P.~Goatin.
\newblock Well-posedness of a conservation law with non-local flux arising in
  traffic flow modeling.
\newblock {\em Numer. Math.}, 132(2):217--241, 2016.

\bibitem{borsche2022PHYSA}
R.~Borsche, A.~Klar, and M.~Zanella.
\newblock Kinetic-controlled hydrodynamics for multilane traffic models.
\newblock {\em Phys. A}, 587:126486, 2022.

\bibitem{chiarello2021CHAPTER}
F.~A. Chiarello.
\newblock An overview of non-local traffic flow models.
\newblock In G.~Puppo and A.~Tosin, editors, {\em Mathematical Descriptions of
  Traffic Flow: Micro, Macro and Kinetic Models}, volume~12 of {\em ICIAM 2019
  SEMA SIMAI Springer Series}, pages 79--91. Springer, 2021.

\bibitem{chiarello2020SIAM}
F.~A. Chiarello, J.~Friedrich, P.~Goatin, and S.~G\"{o}ttlich.
\newblock Micro-{M}acro limit of a non-local generalized {A}w-{R}ascle type
  model.
\newblock {\em SIAM J. Appl. Math.}, 80(4):1841--1861, 2020.

\bibitem{chiarello2020EJAM}
F.~A. Chiarello, J.~Friedrich, P.~Goatin, S.~G{\"o}ttlich, and O.~Kolb.
\newblock A non-local traffic flow model for 1-to-1 junctions.
\newblock {\em Eur. J. Appl. Math.}, 31(6):1029--1049, 2020.

\bibitem{chiarello2018M2AN}
F.~A. Chiarello and P.~Goatin.
\newblock Global entropy weak solutions for general non-local traffic flow
  models with anisotropic kernel.
\newblock {\em ESAIM Math. Model. Numer. Anal.}, 52(1):163--180, 2018.

\bibitem{chiarello2019NHM}
F.~A. Chiarello and P.~Goatin.
\newblock Non-local multi-class traffic flow models.
\newblock {\em Netw. Heterog. Media}, 14(2):371--387, 2019.

\bibitem{chiarello2021MMS}
F.~A. Chiarello, B.~Piccoli, and A.~Tosin.
\newblock Multiscale control of generic second order traffic models by
  driver-assist vehicles.
\newblock {\em Multiscale Model. Simul.}, 19(2):589--611, 2021.

\bibitem{chiarello2021IJNM}
F.~A. Chiarello, B.~Piccoli, and A.~Tosin.
\newblock A statistical mechanics approach to macroscopic limits of
  car-following traffic dynamics.
\newblock {\em Internat. J. Non-Linear Mech.}, 137:103806/1--11, 2021.

\bibitem{daganzo1995TR}
C.~F. Daganzo.
\newblock Requiem for second-order fluid approximation of traffic flow.
\newblock {\em Transportation Res.}, 29(4):277--286, 1995.

\bibitem{difrancesco2017MBE}
M.~Di~Francesco, S.~Fagioli, and M.~Rosini.
\newblock Many particle approximation of the {A}w-{R}ascle-{Z}hang second order
  model for vehicular traffic.
\newblock {\em Math. Biosci. Eng.}, 14(1):127--141, 2017.

\bibitem{difrancesco2015ARMA}
M.~Di~Francesco and M.~D. Rosini.
\newblock Rigorous derivation of nonlinear scalar conservation laws from
  {F}ollow-the-{L}eader type models via many particle limit.
\newblock {\em Arch. Ration. Mech. Anal.}, 217(3):831--871, 2015.

\bibitem{dimarco2020JSP}
G.~Dimarco and A.~Tosin.
\newblock The {A}w-{R}ascle traffic model: {E}nskog-type kinetic derivation and
  generalisations.
\newblock {\em J. Stat. Phys.}, 178(1):178--210, 2020.

\bibitem{dimarco2022JSP}
G.~Dimarco, A.~Tosin, and M.~Zanella.
\newblock Kinetic derivation of {A}w--{R}ascle--{Z}hang-type traffic models
  with driver-assist vehicles.
\newblock {\em J. Stat. Phys.}, 186(1):17/1--26, 2022.

\bibitem{fornasier2011PHYSD}
M.~Fornasier, J.~Haskovec, and G.~Toscani.
\newblock Fluid dynamic description of flocking via the {P}ovzner-{B}oltzmann
  equation.
\newblock {\em Phys. D}, 240(1):21--31, 2011.

\bibitem{fraia2020RUMI}
M.~Fraia and A.~Tosin.
\newblock The {B}oltzmann legacy revisited: kinetic models of social
  interactions.
\newblock {\em Mat. Cult. Soc. Riv. Unione Mat. Ital. (I)}, 5(2):93--109, 2020.

\bibitem{friedrich2018NHM}
J.~Friedrich, O.~Kolb, and S.~G\"{o}ttlich.
\newblock A {G}odunov type scheme for a class of {LWR} traffic flow models with
  non-local flux.
\newblock {\em Netw. Heterog. Media}, 13(4):531--547, 2018.

\bibitem{gazis1961OR}
D.~C. Gazis, R.~Herman, and R.~W. Rothery.
\newblock Nonlinear follow-the-leader models of traffic flow.
\newblock {\em Oper. Res.}, 9:545--567, 1961.

\bibitem{goatin2017CMS}
P.~Goatin and F.~Rossi.
\newblock A traffic flow model with non-smooth metric interaction:
  well-posedness and micro-macro limit.
\newblock {\em Commun. Math. Sci.}, 15(1):261--287, 2017.

\bibitem{goatin2016NHM}
P.~Goatin and S.~Scialanga.
\newblock Well-posedness and finite volume approximations of the {LWR} traffic
  flow model with non-local velocity.
\newblock {\em Netw. Heterog. Media}, 11(1):107--121, 2016.

\bibitem{herty2003SISC}
M.~Herty and A.~Klar.
\newblock Modeling, simulation, and optimization of traffic flow networks.
\newblock {\em SIAM J. Sci. Comput.}, 25(3):1066--1087, 2003.

\bibitem{herty2010KRM}
M.~Herty and L.~Pareschi.
\newblock {F}okker-{P}lanck asymptotics for traffic flow models.
\newblock {\em Kinet. Relat. Models}, 3(1):165--179, 2010.

\bibitem{herty2007NHM}
M.~Herty, L.~Pareschi, and M.~Sea\"{i}d.
\newblock Enskog-like discrete velocity models for vehicular traffic flow.
\newblock {\em Netw. Heterog. Media}, 2(3):481--496, 2007.

\bibitem{herty2020KRM}
M.~Herty, G.~Puppo, S.~Roncoroni, and G.~Visconti.
\newblock The {BGK} approximation of kinetic models for traffic.
\newblock {\em Kinet. Relat. Models}, 13(2):279--307, 2020.

\bibitem{herty2021SEMA-SIMAI}
M.~Herty, A.~Tosin, G.~Visconti, and M.~Zanella.
\newblock Reconstruction of traffic speed distributions from kinetic models
  with uncertainties.
\newblock In G.~Puppo and A.~Tosin, editors, {\em Mathematical Descriptions of
  Traffic Flow: Micro, Macro and Kinetic Models}, volume~12 of {\em ICIAM 2019
  SEMA SIMAI Springer Series}, pages 1--16. Springer, 2021.

\bibitem{keimer2017JDE}
A.~Keimer and L.~Pflug.
\newblock Existence, uniqueness and regularity results on nonlocal balance
  laws.
\newblock {\em J. Differential Equations}, 263(7):4023--4069, 2017.

\bibitem{klar1997JSP}
A.~Klar and R.~Wegener.
\newblock Enskog-like kinetic models for vehicular traffic.
\newblock {\em J. Stat. Phys.}, 87(1-2):91--114, 1997.

\bibitem{klar2000SIAP}
A.~Klar and R.~Wegener.
\newblock Kinetic derivation of macroscopic anticipation models for vehicular
  traffic.
\newblock {\em SIAM J. Appl. Math.}, 60(5):1749--1766, 2000.

\bibitem{lachowicz1990ARMA}
M.~Lachowicz and M.~Pulvirenti.
\newblock A stochastic system of particles modelling the {E}uler equations.
\newblock {\em Arch. Ration. Mech. Anal.}, 109(1):81--93, 1990.

\bibitem{lighthill1955PRSLA}
M.~J. Lighthill and G.~B. Whitham.
\newblock On kinematic waves. {II}. {A} theory of traffic flow on long crowded
  roads.
\newblock {\em Proc. R. Soc. Lond. A}, 229(1178):317--345, 1955.

\bibitem{pareschi2013BOOK}
L.~Pareschi and G.~Toscani.
\newblock {\em Interacting {M}ultiagent {S}ystems: {K}inetic equations and
  {M}onte {C}arlo methods}.
\newblock Oxford University Press, 2013.

\bibitem{payne1971MMPS}
H.~J. Payne.
\newblock Models of freeway traffic and control.
\newblock {\em Math. Models Publ. Sys.}, 28:51--61, 1971.

\bibitem{piccoli2020ZAMP}
B.~Piccoli, A.~Tosin, and M.~Zanella.
\newblock Model-based assessment of the impact of driver-assist vehicles using
  kinetic theory.
\newblock {\em Z. Angew. Math. Phys.}, 71(5):152/1--25, 2020.

\bibitem{povzner1962AMSTS}
A.~Y. Povzner.
\newblock The {B}oltzmann equation in kinetic theory of gases.
\newblock {\em Amer. Math. Soc. Transl. Ser. 2}, 47:193--216, 1962.

\bibitem{prigogine1960OR}
I.~Prigogine and F.~C. Andrews.
\newblock A {B}oltzmann-like approach for traffic flow.
\newblock {\em Operations Res.}, 8(6):789--797, 1960.

\bibitem{prigogine1971BOOK}
I.~Prigogine and R.~Herman.
\newblock {\em Kinetic theory of vehicular traffic}.
\newblock American Elsevier Publishing Co., New York, 1971.

\bibitem{rascle2002MCM}
M.~Rascle.
\newblock An improved macroscopic model of traffic flow: derivation and links
  with the {L}ighthill-{W}hitham model.
\newblock {\em Math. Comput. Modelling}, 35(5--6):581--590, 2002.

\bibitem{richards1956OR}
P.~I. Richards.
\newblock Shock waves on the highway.
\newblock {\em Operations Res.}, 4:42--51, 1956.

\bibitem{sopasakis2006SIAP}
A.~Sopasakis and M.~A. Katsoulakis.
\newblock Stochastic modeling and simulation of traffic flow: asymmetric single
  exclusion process with {A}rrhenius look-ahead dynamics.
\newblock {\em SIAM J. Appl. Math.}, 66(3):921--944, 2006.

\bibitem{tosin2019MMS}
A.~Tosin and M.~Zanella.
\newblock Kinetic-controlled hydrodynamics for traffic models with
  driver-assist vehicles.
\newblock {\em Multiscale Model. Simul.}, 17(2):716--749, 2019.

\bibitem{tosin2021SEMA-SIMAI}
A.~Tosin and M.~Zanella.
\newblock {B}oltzmann-type description with cutoff of {F}ollow-the-{L}eader
  traffic models.
\newblock In G.~Albi, S.~Merino-Aceituno, A.~Nota, and M.~Zanella, editors,
  {\em Trails in Kinetic Theory: Foundational Aspects and Numerical Methods},
  volume~25 of {\em SEMA SIMAI Springer Series}, pages 227--251. Springer,
  2021.

\bibitem{tosin2021MCMF}
A.~Tosin and M.~Zanella.
\newblock Uncertainty damping in kinetic traffic models by driver-assist
  controls.
\newblock {\em Math. Control Relat. Fields}, 11(3):681--713, 2021.

\bibitem{zhang2002TRB}
H.~M. Zhang.
\newblock A non-equilibrium traffic model devoid of gas-like behavior.
\newblock {\em Transportation Res. Part B}, 36(3):275--290, 2002.

\end{thebibliography}
\end{document}